\documentclass[11pt]{article}

\usepackage[top=1.0in,right=1.0in,left=1.0in,bottom=1.75in]{geometry}
\usepackage{amssymb,amsbsy}
\usepackage{amsmath,amscd,amsfonts}
\usepackage{amsthm}
\usepackage{graphics,graphicx,epstopdf}
\usepackage[T1]{fontenc}
\usepackage[svgnames,dvipsnames,table]{xcolor}
\usepackage{subfig}
\usepackage[affil-it,auth-sc]{authblk}
\usepackage{stmaryrd}
\usepackage[svgnames,dvipsnames,table]{xcolor}
\usepackage{pgfplots}
\usepackage[siunitx]{circuitikz}
\usepackage{booktabs}
\usepackage{caption}
\usepackage{multicol}
\usepackage[]{jabbrv}
\usepackage[numbers,sort&compress]{natbib}
\usepackage[utf8]{inputenc}
\usepackage[english]{babel}
\usepackage{braket}
\usepackage{mathtools}
\usepackage{etoolbox}
\usepackage[referable]{threeparttablex}
\appto\TPTnoteSettings{\footnotesize}
\setTableNoteFont{\footnotesize}
\usepackage{multirow}
\usepackage{rotating}
\usepackage{needspace} 
\usepackage[section]{placeins} 
\usepackage{enumerate}
\usepackage[breaklinks]{hyperref}
\hypersetup{
    colorlinks,
    linkcolor={blue!80!black},
    citecolor={blue!80!black},
    urlcolor={blue!80!black}
}
\newcommand{\beq}{\begin{equation}}
\newcommand{\eeq}{\end{equation}}

\newcommand{\beqar}{\begin{eqnarray}}
\newcommand{\eeqar}{\end{eqnarray}}
\newcommand{\bit}{\begin{itemize}}
\newcommand{\eit}{\end{itemize}}
\newcommand{\benum}{\begin{enumerate}}
\newcommand{\eenum}{\end{enumerate}}
\newcommand{\barr}{\begin{array}}
\newcommand{\earr}{\end{array}}

\def\scalp{\mbox{\boldmath $\, \cdot \,$}}

\def\XXint#1#2#3{{\setbox0=\hbox{$#1{#2#3}{\int}$}
   \vcenter{\hbox{$#2#3$}}\kern-.5\wd0}}

\def\b0{\mbox{\boldmath $0$}}

\def\bI{\mbox{\boldmath $I$}}

\def\bQ{\mbox{\boldmath $Q$}}

\def\Id{\mbox{\boldmath $I$}}
\newcommand{\bsigma}{\mbox{\boldmath $\sigma$}}

\newcommand{\bvarepsilon}{\mbox{\boldmath $\varepsilon$}}

\def\f0{\ensuremath{\mathbb{O}}}

\newcommand{\tr}{\mathop{\mathrm{tr}}}
\newcommand{\dev}{\mathop{\mathrm{dev}}}

\newcolumntype{L}{>{$}l<{$}}
\newcolumntype{C}{>{$}c<{$}}
\newcolumntype{R}{>{$}r<{$}}
\newcommand{\av}[1]{\langle #1 \rangle}
\definecolor{Gray}{gray}{0.9}

\graphicspath{{./}{./figures/}}
\definecolor{Light-Green}{RGB}{51, 255, 51}
\definecolor{Light-Red}{RGB}{255, 55, 55}

\title{Computational modelling and experimental validation \\ of industrial forming processes \\ by cold pressing of aluminum silicate powder}
\author[1]{M. Penasa}
\author[2]{L. Argani}
\author[1]{D. Misseroni}
\author[1]{F. Dal Corso}
\author[3]{\\ M. Cova}
\author[1]{A. Piccolroaz\footnote{Corresponding author: e-mail: roaz@ing.unitn.it; phone: +39\,0461\,282583.}}

\affil[1]{Dipartimento di Ingegneria Meccanica e Strutturale, Universit\`a di Trento, Italy}
\affil[2]{Department of Mathematical Sciences, University of Liverpool, UK}
\affil[3]{Ceramics \& Tiles Technical Office, Sacmi Imola S.C., Italy}

\date{}

\begin{document}

\maketitle

\vspace{-3mm}

\begin{abstract}
\noindent

A novel approach to the modelling and simulation of the industrial compaction process of ceramic powders is proposed, based on a combination of: (i) continuum mechanics modelling of the constitutive response of the material; (ii) finite element discretization and computer implementation of the mechanical model; (iii) parametric identification by a multi-objective optimization of simulated experimental tests. 
The capabilities of the proposed approach are highlighted through computer simulations of realistic industrial compaction processes, namely, the forming of an axisymmetric tablet and of a three-dimensional ceramic tile. 

The presented methods and the pointed out results pave the way for the introduction of so-called virtual prototyping into the industrial practice of ceramic forming processes.

\end{abstract}



%

\section{Introduction}
\label{sec01}

Ceramic forming by cold pressing of powders is a common practice in both traditional and advanced ceramic technologies. Due to its industrial interest, this process has been the focus of much attention by the research community over the past decades. Two main approaches in the modelling of granular matter can be distinguished: a micromechanical approach, to analyse the deformation of individual granules in detail, and a continuum macroscopic approach, to describe averaged deformations at the macroscale.

The micromechanical approach has led to the development of the discrete element method (closely related to molecular dynamics) \citep{Cundall1979,Cundall1988,Hart1988,Nosewicz2013,Rojek2013175}. This method can accurately describe the granular flow during the first stage of the compaction process (low pressure), but it is excessively detailed for later stages (high pressure) where the material is better described as a porous solid \citep{Thornton1997,Jerier2011}. Moreover, this method has the disadvantage that the size of the sample (number of particles) and the duration of a simulation are limited by the available computational power \citep{OSullivan2006,Thakur2015}.
For this reason, the continuum macroscopic approach is preferable for the simulation of forming of large components. Moreover, the continuum material model can be implemented in a finite element computer code, which is more accessible to the industry than the discrete element method.

Based on the macroscopic approach, a new rational strategy is proposed in the present article for the computer simulation of real, large-scale ceramic component forming. The aim is to provide the ceramic manufacturers with an effective tool for the optimal design of moulds, punches, and presses for ceramic forming and, thus, to make virtual prototyping a realistic technology for the ceramic industry.

The paper is organized as follows. The continuum mechanics approach to ceramic powder densification is presented in Sec.~\ref{sec03}. This includes the theoretical model as well as its finite element implementation. In Sec.~\ref{sec04} the technique for parameter identification is described by a multi-objective optimization of simulations of experimental tests. Finally, in Sec.~\ref{sec05}, numerical simulations are presented of industrial powder compaction processes, namely axisymmetric tablet formation and three-dimensional tile forming. Die wall friction and deformation of the mould are taken into account. Simulated densities and lateral forces on die walls are compared with experimental results.

\section{Material}
\label{sec02}

The material considered in the present study is the aluminum silicate spray dried powder manufactured by Sacmi S.C.\ (Imola, Italy), labelled I14730, and described in \citet{Bosi2013}. Two different water contents are considered, namely, $w=5.5\%$ and $w=7.5\%$, corresponding to  values used in the industrial forming of traditional ceramics. The granule density, obtained with an helium pycnometer \citep{Dellavolpe2006}, is $\rho_t = 2.599$ g/cm$^3$.

\section{Constitutive framework for ceramic powder densification}
\label{sec03}

\subsection{Constitutive model}

The constitutive model was originally developed for the description of the compaction of alumina powder by Bigoni and co-workers \citep{Piccolroaz2006p1,Piccolroaz2006p2,Stupkiewicz2014b,Penasa2014,Stupkiewicz2015}. That model has been modified in the present work, so as to make it more suitable for the description of the compaction of aluminum silicate powder. 

All the essential equations of the constitutive model are given in Box 1. For details, the interested reader is referred to the above-mentioned references. The main features of the model can be summarized as follows:
\begin{enumerate}[(a)]
 
 \item Non-linear elastic law
 
 Granular materials typically show non-linear response in the elastic regime, whereas partially and fully densified green bodies behave linearly. The elastic law adopted in the model, Eq.~(\ref{elas}), is able to describe this transition of elastic properties.
 
 \item Extremely flexible yield function \citep[so-called \lq BP yield function',][]{Bigoni2004}
 
 Powders and dense materials are described by yield loci of remarkably different shape. The BP yield function has the unique feature  to continuously describe a transition between yield surfaces typical of different materials.
 
 \item Cooper-Eaton hardening law \citep{Cooper1962}
 
 The first hardening law, Eq.~(\ref{hard1}), describes the densification behaviour of the material subject to isotropic compression. This law is based on a micro-mechanical model originally proposed by \citet{Cooper1962}.
 
 \item Increase in cohesion
 
 The compaction process of a ceramic powder gives a form to the green body, making it at the same time cohesive and tractable for subsequent processing. The second hardening law, Eq.~(\ref{hard2}), describes the increase in cohesion at increasing forming pressure.
 
 \item Elasto-plastic coupling
 
 During the compaction of a ceramic powder, the elastic stiffness of the material increases, a phenomenon which is clearly visible, for instance, from the unloading curves of a simple uniaxial compaction test performed at different final forming pressures \citep[see][]{Stupkiewicz2014b}. The elastic stiffening of a ceramic powder during densification is connected with the volumetric plastic deformation of the material and has been addressed both experimentally and computationally \citep{Argani2015arxiv1,Argani2015arxiv2}. It is accounted for by incorporating into the model the so-called \lq elasto-plastic coupling' \citep{Hueckel1975,Hueckel1976,Dougill1976,Bigoni2012}. For a description of the concept of elasto-plastic coupling and its use to model ceramic powder compaction, see for instance \citet{Stupkiewicz2014a}.

\end{enumerate}

Compression and extension triaxial tests have been performed on aluminum silicate powder, commonly used in industrial practice to produce ceramic tiles \citep[see][]{Bosi2013}. These experimental results, for the first time available for aluminum silicate, indicated the necessity to introduce into the modelling a new hardening law describing the increase of deviatoric strength as related to the deviatoric plastic deformation, see Eq.~(\ref{hard3}) in Box 1. 

This hardening law has been adapted from the class of isotropic hardening laws proposed in \citet{Poltronieri2014} to describe the nonlinear behaviour of concrete. These hardening laws display two crucial features: (i) they can be given both in an incremental and in the corresponding finite form; (ii) they describe a smooth transition from linear elastic to plastic behaviour, incorporating linear and nonlinear hardening, and may approach the perfectly plastic limit in the latter case. In particular, all the three hardening laws adopted in the present model can be formulated in a finite form, see Eqs.~(\ref{hard1})--(\ref{hard3}) in Box 1, which allows for a more efficient finite element implementation. 

The constitutive model is defined by \num{22} material parameters (see Tab.~\ref{tab:alluminium-silicate-parameters}), so that an identification of these material parameters is needed and will be performed through a technique combining direct fitting of experimental results together with a multi-objective optimization on simulated experiments (Sec.~\ref{sec04}).

\begin{center}

\framebox{
\begin{minipage}{0.9\columnwidth}

\paragraph{Box 1: Constitutive equations of the model for ceramic powder compaction}

\begin{enumerate}

\item Additive split of strain into an elastic $\bvarepsilon^e$ and plastic $\bvarepsilon^p$ components:
\begin{equation}
\bvarepsilon = \bvarepsilon^e + \bvarepsilon^p
\end{equation}

\item Non-linear elastic stress/strain law:
\begin{multline}
\label{elas}
\bsigma(\bvarepsilon_e, e_v^p) =
\left\{ -\frac{2}{3} \mu\, e_v^e + c \right. \\
\left. + (p_0 + c)
\left[ \left(d(e_v^p) - \frac{1}{d(e_v^p)}\right) \frac{(1 + e_0)e_v^e}{\kappa}
- \exp \left( -\frac{(1 + e_0)e_v^e}{d(e_v^p)^{1/n} \kappa} \right)
\right] \right\} \Id + 2 \mu\, \bvarepsilon^{e}, 
\end{multline}
where $\bsigma$ is the stress, $e_v^e = \tr \bvarepsilon^e$ and $e_v^p = \tr \bvarepsilon^p$ are the elastic and plastic volumetric strains, respectively, $p_0$ is the initial confinement and $e_0$ the initial void ratio.

\item Elasto-plastic coupling:
\begin{equation}
\label{epcoupling}
d = 1 + B \av{p_c - p_{cb}}, \quad \mu(d) = \mu_0 + c\left(d - \frac{1}{d}\right) \mu_1,
\end{equation}

\item BP yield function:
\begin{equation}
F(\bsigma, M, p_c,c) = f(p, M, p_c, c) + q\, g(\theta),
\end{equation}
where $p=\tr\bsigma$, $q=\sqrt{3\dev\bsigma\scalp\dev\bsigma/2}$, and $f(p)$ and $g(\theta)$ are the meridian and deviatoric functions:
\begin{equation}
\label{meridian}
f(p, M, p_c, c) =
-M p_c \sqrt{\left[\phi - \phi^m\right]\left[2 (1 - \alpha) \phi + \alpha\right]}, \quad \phi = \frac{p + c}{p_c + c},
\end{equation}
\begin{equation}
\label{deviatoric}
g(\theta) = \cos{\left[ \beta \frac{\pi}{6} - \frac{1}{3} \cos^{-1} \left(\gamma \cos{3 \theta}\right)\right]}.
\end{equation}

\item Non-associative plastic flow rule:
\begin{equation}
\label{nonassoc}
\dot\bvarepsilon^p = \dot\lambda \left[\bQ - \frac{1}{3}\epsilon(1 - \phi) (\tr\bQ) \bI\right], \quad \bQ = \frac{\partial F}{\partial \bsigma}.
\end{equation}

\item Hardening laws:
\begin{align}
\label{hard1}
e_v^p &= -\frac{e_0}{1 + e_0}\left\{a_1 \exp{\left(-\frac{\varLambda_1}{p_c}\right)} + a_2 \exp{\left(-\frac{\varLambda_2}{p_c}\right)}\right\}, \\[3mm]
\label{hard2}
c &= c_{\infty} \left[ 1- \exp \left(-\varGamma <p_c - p_{cb}> \right) \right], \\[3mm]
\label{hard3}
M &= M_0 + \frac{k_1}{\delta} \frac{(1 + \delta J_2^p)^{n-1} - 1}{(n-1)(1 + \delta J_2^p)^{n-1}}, \quad 
J_2 = \frac{1}{2} \dev \bvarepsilon^p \cdot \bvarepsilon^p.
\end{align}
\end{enumerate}

\end{minipage}
}

\end{center}

\subsection{Finite Element implementation and integration of the material model into commercial FEM codes}

The most efficient way to implement an elastoplastic constitutive model, to be used with commercial Finite Element software, is to develop an external subroutine describing the material response. 

This procedure can be carried out in Abaqus FEA by coding a UMAT (User MATerial) subroutine, which interfaces with the FE software through a standardized parameter list in the subroutine call statement. The UMAT subroutine is compiled and linked to the main Abaqus executable prior to the job execution.

The implementation of the constitutive model for ceramic powder compaction sketched in Box 1  must overcome non-standard difficulties, which include nonlinear elastic behavior, even at small strain, and elastoplastic coupling. Furthermore, the ‘stretchable’ pressure-sensitive yield function introduced by \citet{Bigoni2004} has the inconvenience that, in order to be convex, must be defined $+ \infty$ in some regions outside the elastic domain. This fact, which prevents the application of standard return-mapping techniques for the solution of the plasticity equations, has been recently overcome by \citet{Brannon2010}, \citet{Penasa2014}, and \citet{Stupkiewicz2014a}, using different strategies. 

The last-mentioned technique, based on a implicit definition of the BP yield function, has been used in the current implementation of the constitutive elastoplastic model. In order to increase the stability and robustness of the subroutine, a fully-implicit return mapping technique has been combined with a substepping procedure \citep{Perez-Foguet2001,Tu2009}.

The development of the UMAT subroutine code\footnote{The developed code has been also translated into a USERMAT subroutine for the use in the Ansys environment.} has been carried out by using the advanced hybrid symbolic-numeric approach implemented in {\em AceGen}  \citep{Korelc2002,Korelc2009}, a symbolic code generator available as a package of {\em Wolfram Mathematica}. The combination of automatic differentiation (AD) technique, optimization of formulae and automatic generation of computer code (such as C++, FORTRAN and Matlab) available in {\em AceGen} made possible to efficiently and rapidly prototype the new numerical procedure and benchmark the generated code, which can also be tested within the Mathematica environment by means of the flexible FE code {\em AceFEM}.


%

\section{Material parameter identification by simulation of experimental tests}
\label{sec04}

\subsection{Experimental tests on aluminum silicate powder}

Some of the material parameters, involved in the constitutive model for ceramic powder densification, were identified directly from the results of a set of experimental tests by \citet{Bosi2013}. In particular, the following tests were performed:
\begin{enumerate}[(a)]
 
 \item Uniaxial deformation test were carried out by imposing compaction of the aluminum silicate powder in a 30 mm diameter mould, filled until a height of 4 mm. After reaching the desired pressure $\sigma_1=\{5,10,30,45,60,80\}$ MPa, the green body was unloaded and extracted from the device.
 
 The uniaxial compaction test provided the force-displacement curves, from which the compaction behaviour of the powder, i.e. the relation between forming pressure and density, was deduced. This allowed the calibration of the first hardening law, Eq.~(\ref{hard1}), and the related material parameters, $a_1$, $\varLambda_1$, $a_2$, $\varLambda_2$. However, the identification of these parameters by a uniaxial compaction test is affected by the fact that the state of stress and deformation is not purely isotropic. This has required an adjustment of the values identified by \citet{Bosi2013}, as explained in Sec.~\ref{sec0404}.
 
 The logarithmic elastic bulk modulus $\kappa$ has been evaluated from the linear elastic phase (at very low pressure, prior to the breakpoint pressure $p_{cb}$). 
 
 \item Equi-biaxial flexure tests were performed on the green body tablets produced in the cylindrical mould, following the ASTM C 1499-05 ‘Standard Test Method for Monotonic Equi-biaxial Flexural Strength of Advanced Ceramics at Ambient Temperature’. This allowed the calibration of the second hardening law, Eq.~(\ref{hard2}), and the related material parameters, $c_\infty$, $\varGamma$, $p_{cb}$, describing the increase of cohesion with forming pressure.
 
 \item Compression and extension triaxial tests have been performed on pre-compacted (at $\sigma_1$ = 40 MPa) cylindrical specimens (38.2 mm diameter and 70 mm height), according to the ASTM D 2664 95a ‘Standard Test Method for Triaxial Compressive Strength of Undrained Rock Core Specimens Without Pore Pressure Measurements’. The results of these tests have been used together with the results of uniaxial compaction tests in the optimization procedure.

\end{enumerate}
All the other material parameters, not identified directly through the mechanical tests, are determined by simulating the experiments and performing an iterative multi-objective optimization, which was carried out with simplified numerical models, involving a small number of finite elements and neglecting the effects of friction. This approach significantly speeded up the FE simulations, so that the entire parameter identification could be performed on a simple laptop computer in a reasonable computational time.
This procedure leads to very accurate results, in particular for the uniaxial deformation tests, which are considered with great interest for effective simulation of industrial tile forming processes.

In both uniaxial deformation and triaxial test simulations, the FE analyses were carried out in the Abaqus FEA environment using axisymmetric 8-node biquadratic elements (CAX8).

\subsection{Simulation of uniaxial deformation tests}

\begin{figure}[!htcb]
\centering
\minipage{0.49\textwidth}
  \includegraphics[width=\linewidth]{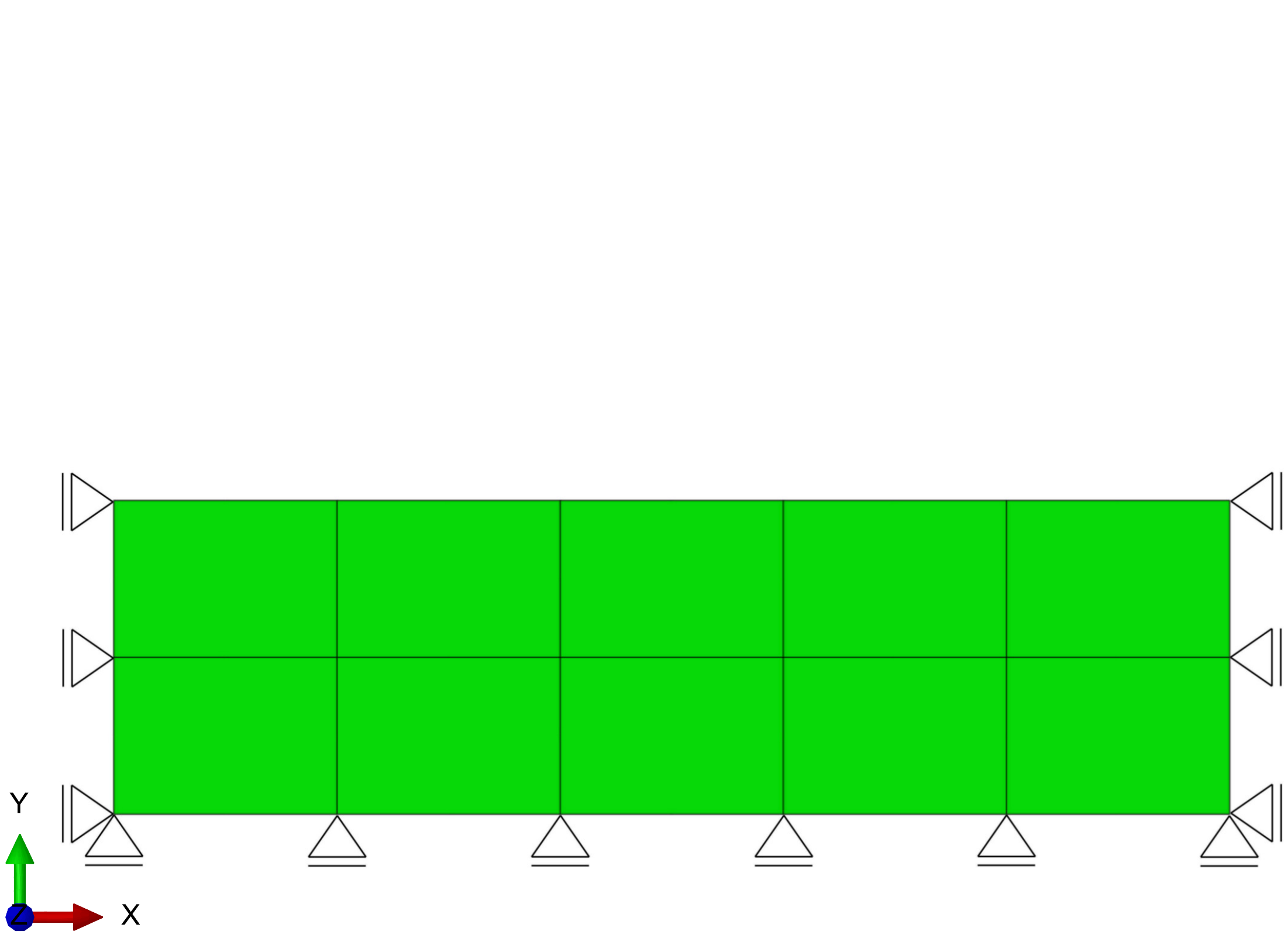}
\endminipage\hfill
\minipage{0.49\textwidth}
  \includegraphics[width=\linewidth]{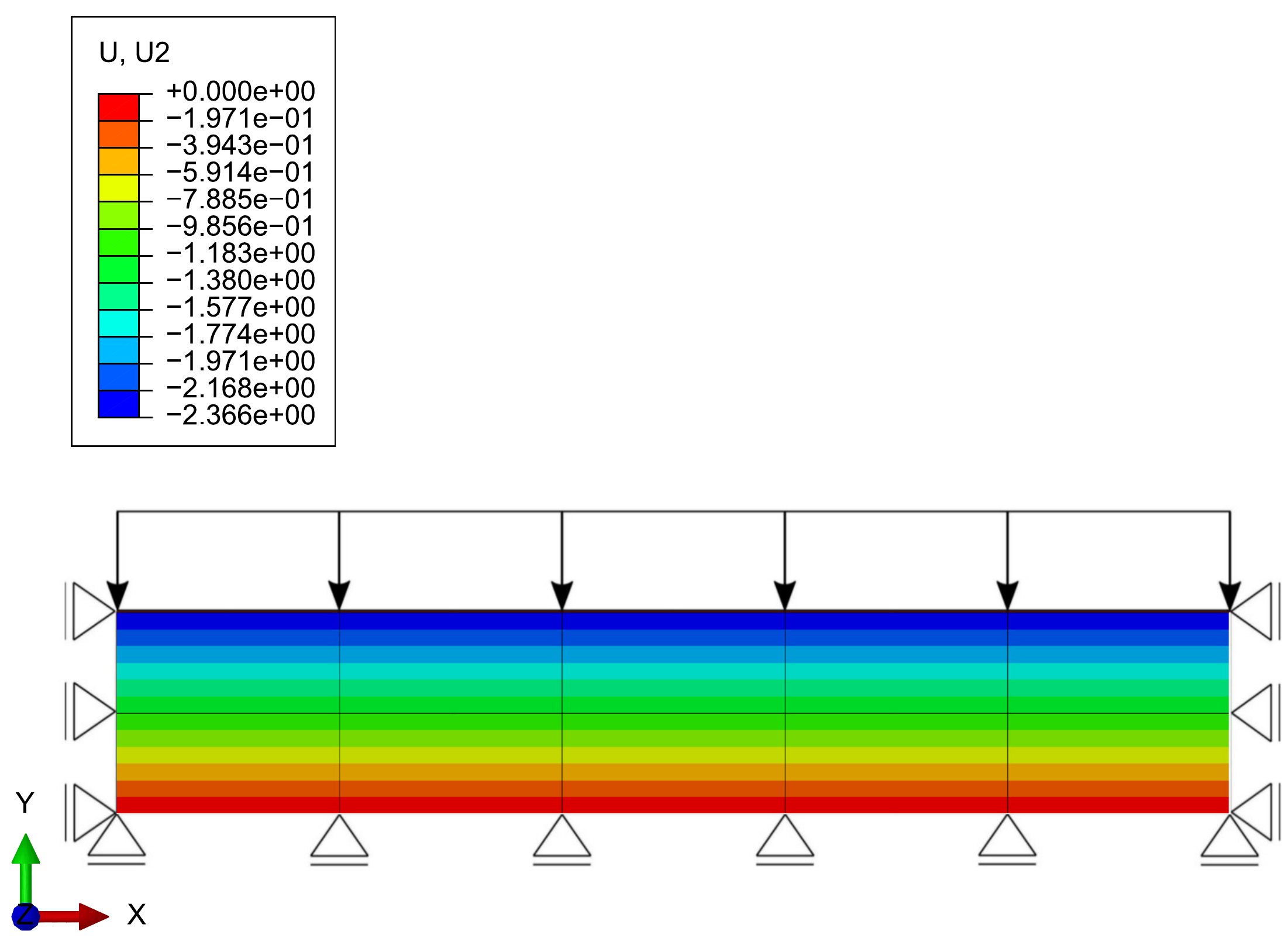}
\endminipage
\caption{\footnotesize Simplified FE model for uniaxial compaction simulation. Undeformed mesh (left) and deformed mesh at the end of loading step (right, contours denote vertical displacement).}
\label{uniaxial-simulation}
\end{figure}

The numerical simulation of uniaxial deformation tests, as performed by \citet{Bosi2013}, involves the execution of the following four steps: 
\begin{enumerate}

\item \label{geostatic} \textbf{Geostatic step}: 
As the powder before compaction is cohesionless, the analysis starts assuming a small  confinement given by an initial value of isotropic stress, $p_0$, which is equilibrated by an equivalent external load in a geostatic step. The initial values of isotropic stress and void ratio used in the simulation are $p_0=0.9$ MPa and $e_0=2.04$.

\item \textbf{Loading}: 
In this step a uniform pressure is applied to the upper face of the sample, see Fig.~\ref{uniaxial-simulation}, while the constraints at the bottom and lateral surfaces reproduce frictionless contact with the mould.

\item \textbf{Unloading}: 
In this phase the pressure on the upper face is removed, while the boundary constraints at bottom and lateral faces are kept active, so as to simulate the removal of the punch.

\item \textbf{Extraction}: 
The supports on the right side of the sample are deactivated in the last step, so that the compact is free to expand transversally, which reproduces ejection from the mould.

\end{enumerate}

Figure~\ref{uniaxial-simulation} shows the undeformed mesh with the constraints to reproduce uniaxial deformation conditions (left) and the deformed mesh at the end of the loading step (right, the contours denote vertical displacement).

\subsection{Simulation of compression and extension triaxial tests}

The FE analysis was carried out to simulate the triaxial tests performed by \citet{Bosi2013}, on cylindrical samples, pre-compacted at 40 MPa. This involves two stages. The first stage corresponds to the preparation of cylindrical specimens, formed by uniaxial compaction at $40$ MPa, followed by an isotropic compaction at $40$ MPa in the triaxial cell. After unloading, the second stage corresponds to the actual triaxial test. Both compression and extension triaxial tests were simulated in the way described below.

\paragraph{Stage 1: forming of cylindrical specimens}

\begin{figure}[!htcb]
\centering
  \includegraphics[width=150mm]{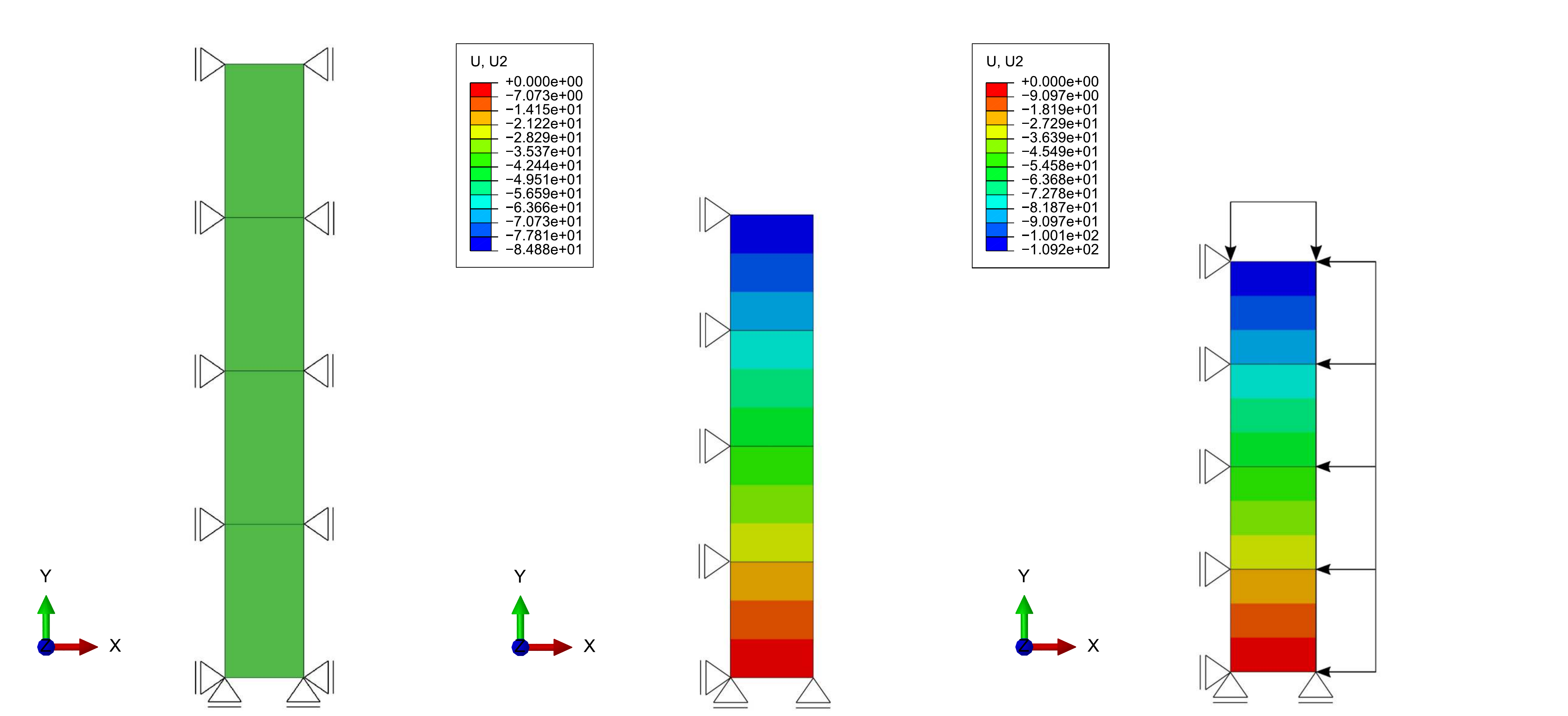}
\caption{\footnotesize Simplified FE model for triaxial test simulation. Undeformed mesh (left), deformed mesh at the end of Stage 1 (forming of cylindrical specimen, centre) and deformed mesh at the end of Stage 2 (triaxial test, right). The contours denote vertical displacement.}
\label{triaxial-simulation}
\end{figure}

\begin{enumerate}

\item \label{geostatic-triaxial} \textbf{Geostatic step}: 
The first geostatic step aims to equilibrate, by imposing an external pressure, $p_0$, the assumed initial isotropic confinement in the ceramic powder. The initial values of isotropic stress and void ratio used in the simulation are $p_0=0.9$ MPa and $e_0=2.04$.

\item \textbf{Uniaxial compaction at $\sigma_2 = 40$ MPa}: The powder is first compacted in a uniaxial deformation step at a final vertical stress equal to 40 MPa. 

\item \textbf{Unloading}: In this step the sample is unloaded.

\item \label{isotropic1} \textbf{Isotropic compaction at $p = 40$ MPa}: 
An uniform pressure of 40 MPa is applied on both faces of the sample (see Fig.~\ref{triaxial-simulation}): $\sigma_1=\sigma_2=40$ MPa.

\item \textbf{Unloading}: In this step the sample is unloaded.

\end{enumerate}

\paragraph{Stage 2: triaxial test}

\begin{enumerate}

\item \textbf{Isotropic loading}: 
The confinement pressure is applied on the faces of the sample: $\sigma_1 = \sigma_2=\{2, 5, 10, 15, 20, 30\}$ MPa.  

\item \textbf{Deviatoric loading}: 
Compression triaxial test: on the upper face of the cylinder a negative displacement is imposed, so as to reduce the hight of the sample. Extension triaxial test: on the lateral face of the cylinder a negative displacement is imposed, so as to reduce the width of the sample.

\end{enumerate}

Figure~\ref{triaxial-simulation} shows the undeformed mesh with the constraints to reproduce uniaxial compaction conditions (step 2 of Stage 1, left), the deformed mesh at the end of forming of the cylindrical specimen (end of Stage 1, centre), the deformed mesh at the end of triaxial test (right). In the central and right figures, contours denote vertical displacement.

\subsection{Material parameter identification by multi-objective optimization}
\label{sec0404}

A multi-objective optimization procedure has been performed in order to identify the constitutive parameters not directly calibrated from the experimental tests performed by \citet{Bosi2013}.

These parameters include: the elasto-plastic coupling parameters, $B$, $n$, $\mu_0$ and $\mu_1$, governing the evolution of elastic properties with plastic deformation, Eq.~(\ref{epcoupling}); the parameters involved in the deviatoric hardening rule, Eq.~(\ref{hard3}), $M_0$, $k_1$, $\delta_1$ and $n_1$; the parameter $\epsilon$ defining the non-associativity, Eq.~(\ref{nonassoc}).

In addition, the parameters $a_1$, $\varLambda_1$, $a_2$, $\varLambda_2$, governing the pressure-density behaviour of the powder in isotropic compression, were included in the optimization. This was done because isotropic compression tests are not available, and thus the values identified by \citet{Bosi2013} through uniaxial compaction tests require an adjustment.

With the introduction of the deviatoric hardening rule, Eq.~(\ref{hard3}), the parameter $M$, describing the pressure-sensitivity, Eq.~(\ref{meridian}), evolves during the densification. This fact is expected to influence also the other parameters governing the meridian shape of the BP yield function, $m$ and $\alpha$, which are thus also included in the optimization.

The multi-objective optimization has been performed by employing the algorithms available in the Dakota Framework \citep{Dakota}, which allows the optimization with both gradient and nongradient-based methods. The chosen optimization strategy aims to find the best possible fit between simulated results and experimental curves, both for uniaxial compression and triaxial compression tests.

The random combination of material parameters, even inside their allowable ranges, may lead to unconsistent results or lack of convergence. For this reason a gradient-free approach has been preferred in the optimization procedure.
As a sufficiently precise starting point was not available, both global and local optimization methods have been used to efficiently estimate the material parameters. 

\begin{figure}[!htcb]
\minipage{0.49\textwidth}
  \includegraphics[width=\linewidth]{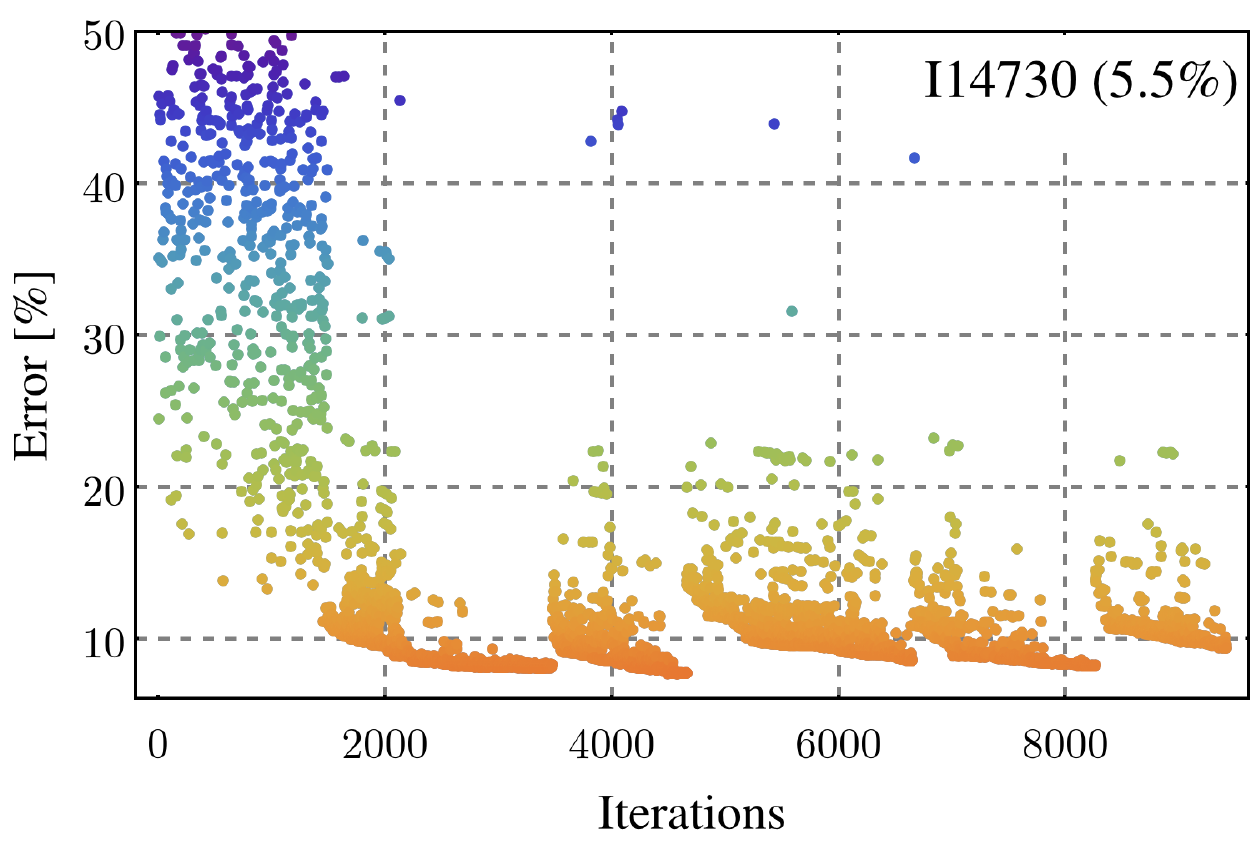}
\endminipage\hfill
\minipage{0.49\textwidth}
  \includegraphics[width=\linewidth]{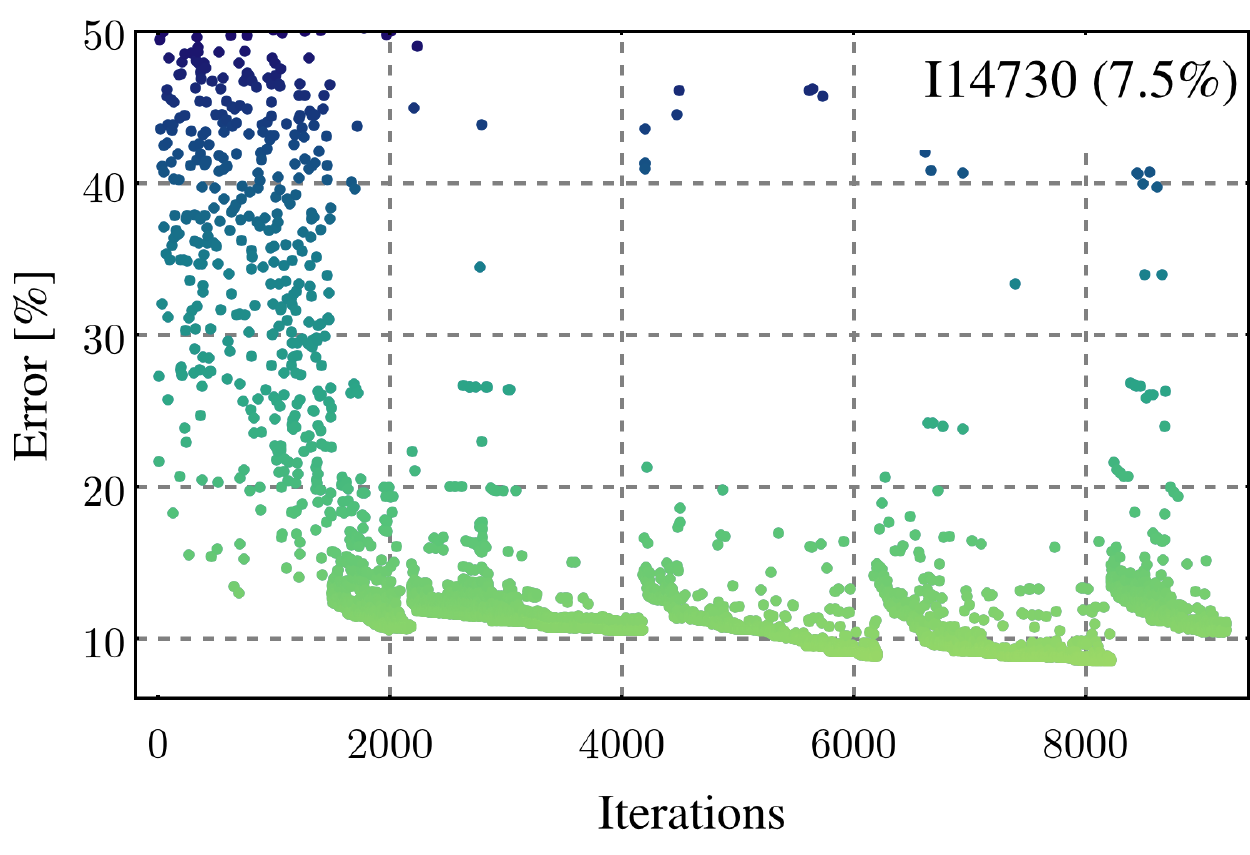}
\endminipage
\caption{\footnotesize Convergence of the hybrid optimization algorithm for the material parameter identification of aluminum silicate powder with $w = 5.5$\% (left) and $w = 7.5$\% (right) water content.}
\label{convergence}
\end{figure}

The ‘hybrid’ procedure involves first a Pareto optimization, by means of moga (multi-objective genetic algorithm). After a sufficiently high number of iterations of the global algorithm, the best five solutions are refined by a local optimization method (pattern-search). The convergence of the optimization strategy, in terms of relative error as a function of the number of iterations, is shown in Fig.~\ref{convergence}. 

The graph shows how, with the increase of the number of iterations, the parameters generating low error tend to become more dense (moga algorithm). After 1500 iterations, the graph shows the convergence of the local optimization algorithm (pattern-search) for the five best solutions of the global algorithm.

In order to reduce the size of the problem, the optimization focused on reaching the best fit of the experimental tests considered of highest industrial interest, namely:
\begin{itemize}
  \item Uniaxial compaction tests at $\sigma_2=$ \{45, 60\} MPa
  \item Triaxial compression tests at cell confinement: $\sigma_1=\sigma_3=$ \{20, 30\} MPa  
\end{itemize}
The final set of parameters for the compaction of aluminum silicate for two different water contents, $w=5.5\%$ and $w=7.5\%$ are reported in Tab. \ref{tab:alluminium-silicate-parameters}.

\begin{table}[!htcb]
\centering
\caption[Material Parameters]{\footnotesize Material parameters for the compaction of aluminum silicate I14730, for two different water contents, $w=5.5\%$ and $w=7.5\%$. Seven parameters were obtained directly from the experiments by \citet{Bosi2013}, the other 15 parameters have been identified by multi-objective optimization.}
\label{tab:alluminium-silicate-parameters}
\begin{tabular}{crLCC}
\toprule
& \multicolumn{2}{c}{\multirow{2}*{Parameter}}	& \multicolumn{2}{c}{\text{Aluminium Silicate I14730}} \\
\cmidrule(lr){4-5}
&	&	& w = \num{5.5}\%	& w = \num{7.5}\% \\
\midrule
\multicolumn{5}{c}{Parameters identified directly from experiments} \\
\midrule
Log. bulk modulus	& (1)	& \kappa	& \num{0.08}	& \num{0.099} \\
\midrule
\multirow{3}*[-2pt]{{Yield surface}}
& (2)	& p_\textup{c0}			& \SI{0.09}{\mega Pa}	& \SI{0.09}{\mega Pa} \\
& (3)	& \beta				& \num{0.1}		& \num{0.08} \\
& (4)	& \gamma			& \num{0.9}		& \num{0.9} \\
\midrule
\multirow{3}*[-2pt]{{Hardening law (\ref{hard2})}}
& (5)	& p_{\textup{cb}}		& \SI{0.22}{\mega Pa}		& \SI{0.17}{\mega Pa} \\
& (6)	& c_{\infty}			& \SI{1.10}{\mega Pa}		& \SI{1.35}{\mega Pa} \\
& (7)	& \Gamma			& \SI{0.06}{\mega Pa^{-1}}	& \SI{0.10}{\mega Pa^{-1}}  \\
\midrule
\multicolumn{5}{c}{Parameters identified by multi-objective optimization} \\
\midrule
\multirow{3}*[-2pt]{{Yield surface}}
& (8)	& M_{0}				& {\num{0.398}}		& {\num{0.506}} \\
& (9)	& m				& {\num{2.26}}		& {\num{3.17}} \\
& (10)	& \alpha			& {\num{1.09}}		& {\num{1.367}} \\
\midrule
\multirow{4}*[-2pt]{{Hardening law (\ref{hard1})}}
& (11)	& a_{1}				& \num{0.763}		& \num{0.780} \\
& (12)	& \Lambda_{1}			& \SI{0.702}{\mega Pa}	& \SI{0.507}{\mega Pa} \\
& (13)	& a_{2}				& \num{0.154}		& \num{0.154} \\
& (14)	& \Lambda_{2} 			& \SI{36.285}{\mega Pa}	& \SI{25.19}{\mega Pa} \\
\midrule
\multirow{3}*[-2pt]{{Hardening law (\ref{hard3})}}
& (15)	& k_{1}				& {\num{301.417}} 	& {\num{30.08}} \\
& (16)	& \delta_{1}			& {\num{456.806}}	& {\num{165.405}} \\
& (17)	& n_{1} 			& {\num{3.647}}		& {\num{9.149}} \\
\midrule
\multirow{4}*[-0.1pt]{{E-P coupling (\ref{epcoupling})}}
& (18)	& B 					& {\SI{9.580}{\mega Pa^{-1}}}		& {\SI{6.908}{\mega Pa^{-1}}} \\
& (19)	& n					& {\num{11.949}}			& {\num{11.749}} \\
& (20)	& \mu_{0}				& {\SI{0.223}{\mega Pa}}		& {\SI{9.822}{\mega Pa}} \\
& (21)	& \mu_{1}				& {\num{24.678}}			& {\num{5.269}}\\
\midrule
Plastic flow (\ref{nonassoc})	& (22)	& \varepsilon		& {\num{0.916}}				& \num{0.586} \\
\bottomrule
\end{tabular}

\end{table}

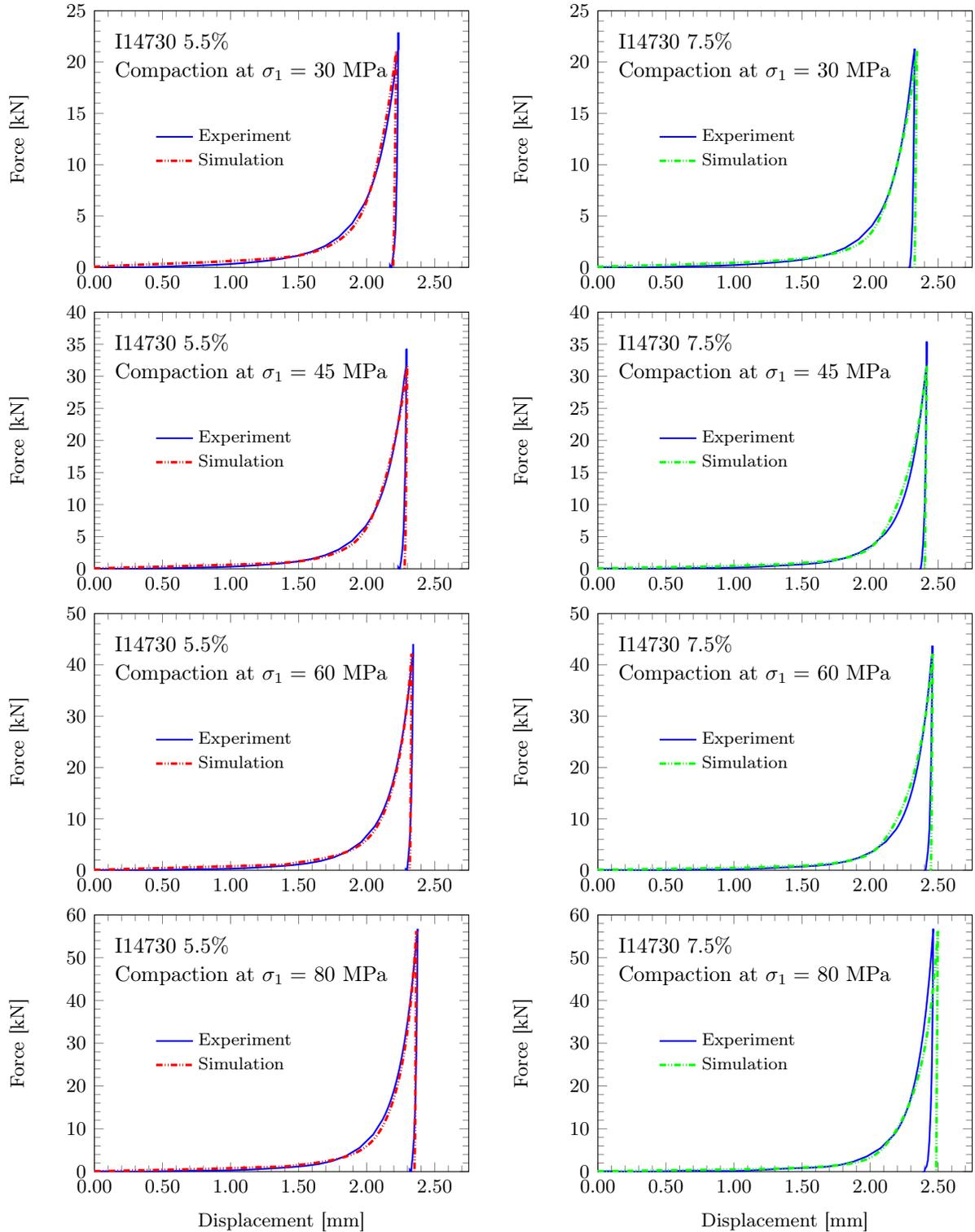
\begin{figure}[!htcb]
\centering
\begin{tikzpicture}

\node[anchor=west] at (0.2,3.7) {\small I14730 5.5\%};
\node[anchor=west] at (0.2,3.2) {\small Compaction at $\sigma_1=$ 30 MPa};

\begin{axis}
[scale only axis,                       
font=\footnotesize,                     
width=0.37\columnwidth,                 
height=0.20\textheight,                 
xmin=0, xmax=2.75, ymin=0, ymax=25,     
scaled ticks=false,                     
   xticklabel style={
     /pgf/number format/precision=2,     
     /pgf/number format/fixed,
     /pgf/number format/fixed zerofill,  
   },
   yticklabel style={
    /pgf/number format/precision=0,     
    /pgf/number format/fixed,
    /pgf/number format/fixed zerofill,  
   },
    xtick={0,0.5,1,1.5,2,2.5}, 
    ytick={0,5,10,15,20,25},  
    minor x tick num = 4,               
    minor y tick num = 4,               
    ylabel={Force [\si{\kilo N}]}, 
  legend style={
  at={(0.35,0.35)},
  anchor=south,
  legend columns=1,
  cells={anchor=west},
  font=\scriptsize,
	draw=none,
}]

\addplot [blue,thick]
file {MCB55-30-experiment.txt};
\addplot [red,densely dashdotdotted, very thick]
file {MCB55-30-simulation.txt};
\legend{Experiment, Simulation}
\end{axis}
\end{tikzpicture} \hfil \begin{tikzpicture}

\node[anchor=west] at (0.2,3.7) {\small I14730 7.5\%};
\node[anchor=west] at (0.2,3.2) {\small Compaction at $\sigma_1=$ 30 MPa};

\begin{axis}
[scale only axis,                       
font=\footnotesize,                     
width=0.37\columnwidth,                 
height=0.20\textheight,                 
xmin=0, xmax=2.75, ymin=0, ymax=25,     
scaled ticks=false,                     
  xticklabel style={
    /pgf/number format/precision=2,     
    /pgf/number format/fixed,
    /pgf/number format/fixed zerofill,  
  },
  yticklabel style={
    /pgf/number format/precision=0,     
    /pgf/number format/fixed,
    /pgf/number format/fixed zerofill,  
  },
    xtick={0,0.5,1,1.5,2,2.5}, 
    ytick={0,5,10,15,20,25},  
    minor x tick num = 4,               
    minor y tick num = 4,               
    ylabel={Force [\si{\kilo N}]}, 
  legend style={
  at={(0.35,0.35)},
  anchor=south,
  legend columns=1,
  cells={anchor=west},
  font=\scriptsize,
	draw=none,
}]

\addplot [blue,thick]
file {MCB75-30-experiment.txt};
\addplot [green,densely dashdotdotted,very thick]
file {MCB75-30-simulation.txt};
\legend{Experiment, Simulation}
\end{axis}
\end{tikzpicture} \\
\begin{tikzpicture}

\node[anchor=west] at (0.2,3.7) {\small I14730 5.5\%};
\node[anchor=west] at (0.2,3.2) {\small Compaction at $\sigma_1=$ 45 MPa};

\begin{axis}
[scale only axis,                       
font=\footnotesize,                     
width=0.37\columnwidth,                 
height=0.20\textheight,                 
xmin=0, xmax=2.75, ymin=0, ymax=40,     
scaled ticks=false,                     
  xticklabel style={
    /pgf/number format/precision=2,     
    /pgf/number format/fixed,
    /pgf/number format/fixed zerofill,  
  },
  yticklabel style={
    /pgf/number format/precision=0,     
    /pgf/number format/fixed,
    /pgf/number format/fixed zerofill,  
  },
    xtick={0,0.5,1,1.5,2,2.5}, 
    ytick={0,5,10,15,20,25,30,35,40},  
    minor x tick num = 4,               
    minor y tick num = 4,               
    ylabel={Force [\si{\kilo N}]}, 
  legend style={
  at={(0.35,0.35)},
  anchor=south,
  legend columns=1,
  cells={anchor=west},
  font=\scriptsize,
	draw=none,
}]

\addplot [blue,thick]
file {MCB55-45-experiment.txt};
\addplot [red,densely dashdotdotted,very thick]
file {MCB55-45-simulation.txt};
\legend{Experiment, Simulation}
\end{axis}
\end{tikzpicture} \hfil \begin{tikzpicture}

\node[anchor=west] at (0.2,3.7) {\small I14730 7.5\%};
\node[anchor=west] at (0.2,3.2) {\small Compaction at $\sigma_1=$ 45 MPa};

\begin{axis}
[scale only axis,                       
font=\footnotesize,                     
width=0.37\columnwidth,                 
height=0.20\textheight,                 
xmin=0, xmax=2.75, ymin=0, ymax=40,     
scaled ticks=false,                     
  xticklabel style={
    /pgf/number format/precision=2,     
    /pgf/number format/fixed,
    /pgf/number format/fixed zerofill,  
  },
  yticklabel style={
    /pgf/number format/precision=0,     
    /pgf/number format/fixed,
    /pgf/number format/fixed zerofill,  
  },
    xtick={0,0.5,1,1.5,2,2.5}, 
    ytick={0,5,10,15,20,25,30,35,40},  
    minor x tick num = 4,               
    minor y tick num = 4,               
    ylabel={Force [\si{\kilo N}]}, 
  legend style={
  at={(0.35,0.35)},
  anchor=south,
  legend columns=1,
  cells={anchor=west},
  font=\scriptsize,
	draw=none,
}]

\addplot [blue,thick]
file {MCB75-45-experiment.txt};
\addplot [green,densely dashdotdotted,very thick]
file {MCB75-45-simulation.txt};
\legend{Experiment, Simulation}
\end{axis}
\end{tikzpicture} \\
\begin{tikzpicture}

\node[anchor=west] at (0.2,3.7) {\small I14730 5.5\%};
\node[anchor=west] at (0.2,3.2) {\small Compaction at $\sigma_1=$ 60 MPa};

\begin{axis}
[scale only axis,                       
font=\footnotesize,                     
width=0.37\columnwidth,                 
height=0.20\textheight,                 
xmin=0, xmax=2.75, ymin=0, ymax=50,     
scaled ticks=false,                     
  xticklabel style={
    /pgf/number format/precision=2,     
    /pgf/number format/fixed,
    /pgf/number format/fixed zerofill,  
  },
  yticklabel style={
    /pgf/number format/precision=0,     
    /pgf/number format/fixed,
    /pgf/number format/fixed zerofill,  
  },
    xtick={0,0.5,1,1.5,2,2.5}, 
    ytick={0,10,20,30,40,50},  
    minor x tick num = 4,               
    minor y tick num = 4,               
    ylabel={Force [\si{\kilo N}]}, 
  legend style={
  at={(0.35,0.35)},
  anchor=south,
  legend columns=1,
  cells={anchor=west},
  font=\scriptsize,
	draw=none,
}]

\addplot [blue,thick]
file {MCB55-60-experiment.txt};
\addplot [red,densely dashdotdotted,very thick]
file {MCB55-60-simulation.txt};
\legend{Experiment, Simulation}
\end{axis}
\end{tikzpicture} \hfil \begin{tikzpicture}

\node[anchor=west] at (0.2,3.7) {\small I14730 7.5\%};
\node[anchor=west] at (0.2,3.2) {\small Compaction at $\sigma_1=$ 60 MPa};

\begin{axis}
[scale only axis,                       
font=\footnotesize,                     
width=0.37\columnwidth,                 
height=0.20\textheight,                 
xmin=0, xmax=2.75, ymin=0, ymax=50,     
scaled ticks=false,                     
  xticklabel style={
    /pgf/number format/precision=2,     
    /pgf/number format/fixed,
    /pgf/number format/fixed zerofill,  
  },
  yticklabel style={
    /pgf/number format/precision=0,     
    /pgf/number format/fixed,
    /pgf/number format/fixed zerofill,  
  },
    xtick={0,0.5,1,1.5,2,2.5}, 
    ytick={0,10,20,30,40,50},  
    minor x tick num = 4,               
    minor y tick num = 4,               
    ylabel={Force [\si{\kilo N}]}, 
  legend style={
  at={(0.35,0.35)},
  anchor=south,
  legend columns=1,
  cells={anchor=west},
  font=\scriptsize,
	draw=none,
}]

\addplot [blue,thick]
file {MCB75-60-experiment.txt};
\addplot [green,densely dashdotdotted,very thick]
file {MCB75-60-simulation.txt};
\legend{Experiment, Simulation}
\end{axis}
\end{tikzpicture} \\
\begin{tikzpicture}

\node[anchor=west] at (0.2,3.7) {\small I14730 5.5\%};
\node[anchor=west] at (0.2,3.2) {\small Compaction at $\sigma_1=$ 80 MPa};

\begin{axis}
[scale only axis,                       
font=\footnotesize,                     
width=0.37\columnwidth,                 
height=0.20\textheight,                 
xmin=0, xmax=2.75, ymin=0, ymax=60,     
scaled ticks=false,                     
  xticklabel style={
    /pgf/number format/precision=2,     
    /pgf/number format/fixed,
    /pgf/number format/fixed zerofill,  
  },
  yticklabel style={
    /pgf/number format/precision=0,     
    /pgf/number format/fixed,
    /pgf/number format/fixed zerofill,  
  },
    xtick={0,0.5,1,1.5,2,2.5}, 
    ytick={0,10,20,30,40,50,60},  
    minor x tick num = 4,               
    minor y tick num = 4,               
    xlabel={Displacement [\si{\milli m}]},
    ylabel={Force [\si{\kilo N}]}, 
  legend style={
  at={(0.35,0.35)},
  anchor=south,
  legend columns=1,
  cells={anchor=west},
  font=\scriptsize,
	draw=none,
}]

\addplot [blue,thick]
file {MCB55-80-experiment.txt};
\addplot [red,densely dashdotdotted,very thick]
file {MCB55-80-simulation.txt};
\legend{Experiment, Simulation}
\end{axis}
\end{tikzpicture} \hfil \begin{tikzpicture}

\node[anchor=west] at (0.2,3.7) {\small I14730 7.5\%};
\node[anchor=west] at (0.2,3.2) {\small Compaction at $\sigma_1=$ 80 MPa};

\begin{axis}
[scale only axis,                       
font=\footnotesize,                     
width=0.37\columnwidth,                 
height=0.20\textheight,                 
xmin=0, xmax=2.75, ymin=0, ymax=60,     
scaled ticks=false,                     
  xticklabel style={
    /pgf/number format/precision=2,     
    /pgf/number format/fixed,
    /pgf/number format/fixed zerofill,  
  },
  yticklabel style={
    /pgf/number format/precision=0,     
    /pgf/number format/fixed,
    /pgf/number format/fixed zerofill,  
  },
    xtick={0,0.5,1,1.5,2,2.5}, 
    ytick={0,10,20,30,40,50,60},  
    minor x tick num = 4,               
    minor y tick num = 4,               
    xlabel={Displacement [\si{\milli m}]},
    ylabel={Force [\si{\kilo N}]}, 
  legend style={
  at={(0.35,0.35)},
  anchor=south,
  legend columns=1,
  cells={anchor=west},
  font=\scriptsize,
	draw=none,
}]

\addplot [blue,thick]
file {MCB75-80-experiment.txt};
\addplot [green,densely dashdotdotted,very thick]
file {MCB75-80-simulation.txt};
\legend{Experiment, Simulation}
\end{axis}
\end{tikzpicture} \\
\caption[]{\footnotesize Uniaxial compaction test: Comparison between experimental results and numerical simulations obtained with the identified material parameters (Tab.~\ref{tab:alluminium-silicate-parameters}), for water content $w = 5.5$\% (left) and $w = 7.5$\% (right). The experimental curves corresponding to forming at $\sigma_2=$ 45 and 60 MPa were used in the optimization procedure.}
\label{fig:MCB}
\end{figure}

\begin{figure}[!htcb]
\centering
\begin{tikzpicture}

\node[anchor=west] at (0.2,3.7) {\small I14730 5.5\%};
\node[anchor=east] at (6.0,0.4) {\small Compression at $\sigma_2=\sigma_3=$ 15 MPa};

\begin{axis}
[scale only axis,                       
font=\footnotesize,                     
width=0.37\columnwidth,                 
height=0.20\textheight,                 
xmin=0, xmax=0.16, ymin=0, ymax=30,     
scaled ticks=false,                     
  xticklabel style={
    /pgf/number format/precision=2,     
    /pgf/number format/fixed,
    /pgf/number format/fixed zerofill,  
  },
  yticklabel style={
    /pgf/number format/precision=0,     
    /pgf/number format/fixed,
    /pgf/number format/fixed zerofill,  
  },
    xtick={0,0.05,0.1,0.15,0.2}, 
    ytick={0,5,10,15,20,25,30},  
    minor x tick num = 4,               
    minor y tick num = 4,               
    ylabel={Von Mises Stress [\si{\mega Pa}]}, 
  legend style={
  at={(0.55,0.25)},
  anchor=south,
  legend columns=1,
  cells={anchor=west},
  font=\scriptsize,
	draw=none,
}]

\addplot [blue,thick]
file {TCB55-15-experiment.txt};
\addplot [red,densely dashdotdotted,very thick]
file {TCB55-15-simulation.txt};
\legend{Experiment, Simulation}
\end{axis}
\end{tikzpicture} \hfil \begin{tikzpicture}

\node[anchor=west] at (0.2,3.7) {\small I14730 7.5\%};
\node[anchor=east] at (6.0,0.4) {\small Compression at $\sigma_2=\sigma_3=$ 15 MPa};

\begin{axis}
[scale only axis,                       
font=\footnotesize,                     
width=0.37\columnwidth,                 
height=0.20\textheight,                 
xmin=0, xmax=0.16, ymin=0, ymax=30,     
scaled ticks=false,                     
  xticklabel style={
    /pgf/number format/precision=2,     
    /pgf/number format/fixed,
    /pgf/number format/fixed zerofill,  
  },
  yticklabel style={
    /pgf/number format/precision=0,     
    /pgf/number format/fixed,
    /pgf/number format/fixed zerofill,  
  },
    xtick={0,0.05,0.1,0.15,0.2}, 
    ytick={0,5,10,15,20,25,30},  
    minor x tick num = 4,               
    minor y tick num = 4,               
    ylabel={Von Mises Stress [\si{\mega Pa}]}, 
  legend style={
  at={(0.55,0.25)},
  anchor=south,
  legend columns=1,
  cells={anchor=west},
  font=\scriptsize,
	draw=none,
}]

\addplot [blue, thick]
file {TCB75-15-experiment.txt};
\addplot [green,densely dashdotdotted,very thick]
file {TCB75-15-simulation.txt};
\legend{Experiment, Simulation}
\end{axis}
\end{tikzpicture} \\
\begin{tikzpicture}

\node[anchor=west] at (0.2,3.7) {\small I14730 5.5\%};
\node[anchor=east] at (6.0,0.4) {\small Compression at $\sigma_2=\sigma_3=$ 0 MPa};

\begin{axis}
[scale only axis,                       
font=\footnotesize,                     
width=0.37\columnwidth,                 
height=0.20\textheight,                 
xmin=0, xmax=0.23, ymin=0, ymax=35,     
scaled ticks=false,                     
  xticklabel style={
    /pgf/number format/precision=2,     
    /pgf/number format/fixed,
    /pgf/number format/fixed zerofill,  
  },
  yticklabel style={
    /pgf/number format/precision=0,     
    /pgf/number format/fixed,
    /pgf/number format/fixed zerofill,  
  },
    xtick={0,0.05,0.1,0.15,0.2}, 
    ytick={0,5,10,15,20,25,30,35},  
    minor x tick num = 4,               
    minor y tick num = 4,               
    ylabel={Von Mises Stress [\si{\mega Pa}]}, 
  legend style={
  at={(0.55,0.25)},
  anchor=south,
  legend columns=1,
  cells={anchor=west},
  font=\scriptsize,
	draw=none,
}]

\addplot [blue,thick]
file {TCB55-20-experiment.txt};
\addplot [red,densely dashdotdotted,very thick]
file {TCB55-20-simulation.txt};
\legend{Experiment, Simulation}
\end{axis}
\end{tikzpicture} \hfil \begin{tikzpicture}

\node[anchor=west] at (0.2,3.7) {\small I14730 7.5\%};
\node[anchor=east] at (6.0,0.4) {\small Compression at $\sigma_2=\sigma_3=$ 20 MPa};

\begin{axis}
[scale only axis,                       
font=\footnotesize,                     
width=0.37\columnwidth,                 
height=0.20\textheight,                 
xmin=0, xmax=0.23, ymin=0, ymax=35,     
scaled ticks=false,                     
  xticklabel style={
    /pgf/number format/precision=2,     
    /pgf/number format/fixed,
    /pgf/number format/fixed zerofill,  
  },
  yticklabel style={
    /pgf/number format/precision=0,     
    /pgf/number format/fixed,
    /pgf/number format/fixed zerofill,  
  },
    xtick={0,0.05,0.1,0.15,0.2}, 
    ytick={0,5,10,15,20,25,30,35},  
    minor x tick num = 4,               
    minor y tick num = 4,               
    ylabel={Von Mises Stress [\si{\mega Pa}]}, 
  legend style={
  at={(0.55,0.25)},
  anchor=south,
  legend columns=1,
  cells={anchor=west},
  font=\scriptsize,
	draw=none,
}]

\addplot [blue,thick]
file {TCB75-20-experiment.txt};
\addplot [green,densely dashdotdotted,very thick]
file {TCB75-20-simulation.txt};
\legend{Experiment, Simulation}
\end{axis}
\end{tikzpicture} \\
\begin{tikzpicture}

\node[anchor=west] at (0.2,3.7) {\small I14730 5.5\%};
\node[anchor=east] at (6.0,0.4) {\small Compression at $\sigma_2=\sigma_3=$ 30 MPa};

\begin{axis}
[scale only axis,                       
font=\footnotesize,                     
width=0.37\columnwidth,                 
height=0.20\textheight,                 
xmin=0, xmax=0.22, ymin=0, ymax=50,     
scaled ticks=false,                     
  xticklabel style={
    /pgf/number format/precision=2,     
    /pgf/number format/fixed,
    /pgf/number format/fixed zerofill,  
  },
  yticklabel style={
    /pgf/number format/precision=0,     
    /pgf/number format/fixed,
    /pgf/number format/fixed zerofill,  
  },
    xtick={0,0.05,0.1,0.15,0.2}, 
    ytick={0,10,20,30,40,50},  
    minor x tick num = 4,               
    minor y tick num = 4,               
    ylabel={Von Mises Stress [\si{\mega Pa}]}, 
  legend style={
  at={(0.55,0.25)},
  anchor=south,
  legend columns=1,
  cells={anchor=west},
  font=\scriptsize,
	draw=none,
}]

\addplot [blue,thick]
file {TCB55-30-experiment.txt};
\addplot [red,densely dashdotdotted,very thick]
file {TCB55-30-simulation.txt};
\legend{Experiment, Simulation}
\end{axis}
\end{tikzpicture} \hfil \begin{tikzpicture}

\node[anchor=west] at (0.2,3.7) {\small I14730 7.5\%};
\node[anchor=east] at (6.0,0.4) {\small Compression at $\sigma_2=\sigma_3=$ 30 MPa};

\begin{axis}
[scale only axis,                       
font=\footnotesize,                     
width=0.37\columnwidth,                 
height=0.20\textheight,                 
xmin=0, xmax=0.22, ymin=0, ymax=50,     
scaled ticks=false,                     
  xticklabel style={
    /pgf/number format/precision=2,     
    /pgf/number format/fixed,
    /pgf/number format/fixed zerofill,  
  },
  yticklabel style={
    /pgf/number format/precision=0,     
    /pgf/number format/fixed,
    /pgf/number format/fixed zerofill,  
  },
    xtick={0,0.05,0.1,0.15,0.2}, 
    ytick={0,10,20,30,40,50},  
    minor x tick num = 4,               
    minor y tick num = 4,               
    ylabel={Von Mises Stress [\si{\mega Pa}]}, 
  legend style={
  at={(0.55,0.25)},
  anchor=south,
  legend columns=1,
  cells={anchor=west},
  font=\scriptsize,
	draw=none,
}]

\addplot [blue,thick]
file {TCB75-30-experiment.txt};
\addplot [green,densely dashdotdotted,very thick]
file {TCB75-30-simulation.txt};
\legend{Experiment, Simulation}
\end{axis}
\end{tikzpicture} \\
\begin{tikzpicture}

\node[anchor=west] at (0.2,3.7) {\small I14730 5.5\%};
\node[anchor=east] at (6.0,0.4) {\small Extension at $\sigma_1=$ 20 MPa};

\begin{axis}
[scale only axis,                       
font=\footnotesize,                     
width=0.37\columnwidth,                 
height=0.20\textheight,                 
xmin=0, xmax=0.025, ymin=0, ymax=18,     
scaled ticks=false,                     
  xticklabel style={
    /pgf/number format/precision=2,     
    /pgf/number format/fixed,
    /pgf/number format/fixed zerofill,  
  },
  yticklabel style={
    /pgf/number format/precision=0,     
    /pgf/number format/fixed,
    /pgf/number format/fixed zerofill,  
  },
    xtick={0,0.01,0.02,0.03}, 
    ytick={0,5,10,15,20,25,30},  
    minor x tick num = 4,               
    minor y tick num = 4,               
    xlabel={Axial Strain},
    ylabel={Von Mises Stress [\si{\mega Pa}]}, 
  legend style={
  at={(0.55,0.25)},
  anchor=south,
  legend columns=1,
  cells={anchor=west},
  font=\scriptsize,
  draw=none,
}]

\addplot [blue,thick]
file {TEB55-20-experiment-smooth.txt};
\addplot [red,densely dashdotdotted,very thick]
file {TEB55-20-simulation.txt};
\legend{Experiment, Simulation}
\end{axis}
\end{tikzpicture} \hfil \begin{tikzpicture}

\node[anchor=west] at (0.2,3.7) {\small I14730 7.5\%};
\node[anchor=east] at (6.0,0.4) {\small Extension at $\sigma_1=$ 30 MPa};

\begin{axis}
[scale only axis,                       
font=\footnotesize,                     
width=0.37\columnwidth,                 
height=0.20\textheight,                 
xmin=0, xmax=0.005, ymin=0, ymax=10.5,     
scaled ticks=false,                     
  xticklabel style={
    /pgf/number format/precision=3,     
    /pgf/number format/fixed,
    /pgf/number format/fixed zerofill,  
  },
  yticklabel style={
    /pgf/number format/precision=0,     
    /pgf/number format/fixed,
    /pgf/number format/fixed zerofill,  
  },
    xtick={0,0.0015,0.003,0.0045}, 
    ytick={0,2,4,6,8,10},  
    minor x tick num = 4,               
    minor y tick num = 4,               
    xlabel={Axial Strain},
    ylabel={Von Mises Stress [\si{\mega Pa}]}, 
  legend style={
  at={(0.65,0.25)},
  anchor=south,
  legend columns=1,
  cells={anchor=west},
  font=\scriptsize,
	draw=none,
}]

\addplot [blue,thick]
file {TEB75-30-experiment-smooth.txt};
\addplot [green,densely dashdotdotted,very thick]
file {TEB75-30-simulation.txt};
\legend{Experiment, Simulation}
\end{axis}
\end{tikzpicture} \\
\caption[]{\footnotesize Compression/extension triaxial test: Comparison between experimental results and numerical simulations obtained with the identified material parameters (Tab.~\ref{tab:alluminium-silicate-parameters}), for water content $w = 5.5$\% (left) and $w = 7.5$\% (right). The experimental curves corresponding to compression tests at confinement pressure $\sigma_1=\sigma_3=$ 20 and 30 MPa were used in the optimization procedure.} 
\label{fig:TCB}
\end{figure}
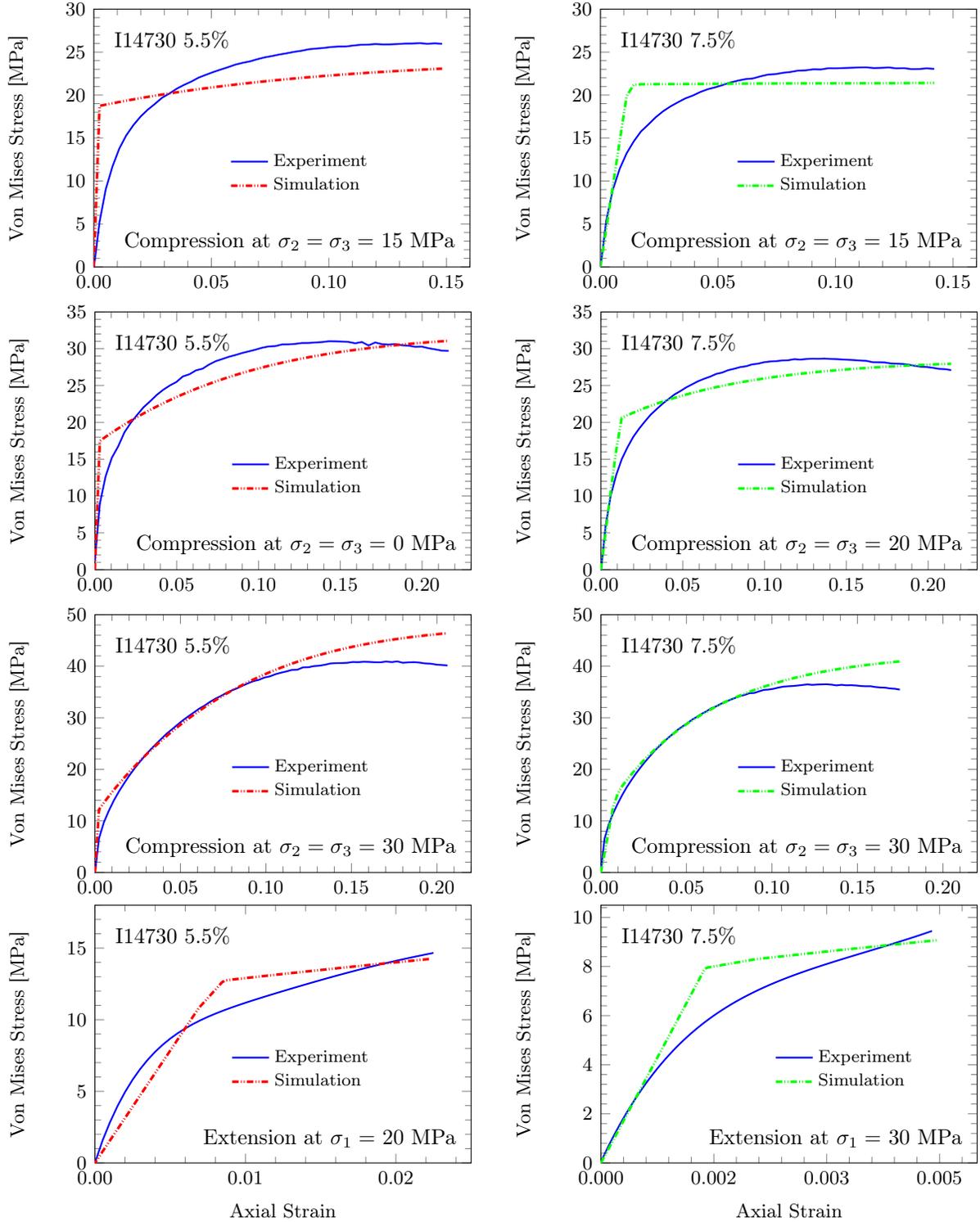

With the identified parameters, it is possible to obtain a good agreement between numerical and experimental results also in tests which are not considered in the optimization. This is an indication of the consistency of the constitutive model and of the validity of the optimization procedure. Figure~\ref{fig:MCB} shows the uniaxial compaction (force vs. displacement) curves for water content $w = 5.5$\% (left) and $w = 7.5$\% (right). Figure~\ref{fig:TCB} shows the compression/extension triaxial (von Mises stress vs.\ axial strain) curves for water content $w = 5.5$\% (left) and $w = 7.5$\% (right).

\section{Numerical simulation of industrial powder compaction processes}
\label{sec05}

\subsection{Experimental identification of friction coefficient between powder and die wall}
\label{friction}

At the end of process of powder compaction, the sample is unloaded and extracted from the forming device. In this phase, friction between the green body and the steel matrix requires the application of an axial force on the tablet to complete the extraction. In the experimental tests performed by \citet{Bosi2013}, it was possible to measure the tangential force $T$ required to remove the green body from the mould. The normal force $N$ on the steel matrix can be calculated from the measured radial deformation of the sample after extraction and the elastic moduli, measured by means of ultrasound technique by \citet{Argani2015arxiv1}. 

The static friction coefficient can be estimated from the relationship between tangential and normal forces acting on the die wall.
The calculation of static friction coefficient for forming pressure of 45 MPa yielded $\mu_s = T/N = $ 0.18. This value has been used in the following FE simulations in order to accurately reproduce the contact interaction between ceramic powder and steel matrix.

\subsection{Numerical simulation of axisymmetric tablet forming}

Finite element analyses, involving contact interaction and friction, have been performed to accurately reproduce the uniaxial compaction tests performed by \citet{Bosi2013}. In these simulations, the complete forming device, composed by matrix, upper and lower punches, was modelled, and the interaction between each single part and the ceramic powder was taken into account.

\begin{figure}[!htcb]
\begin{center}
\minipage{0.30\textwidth}
  \includegraphics[width=\linewidth]{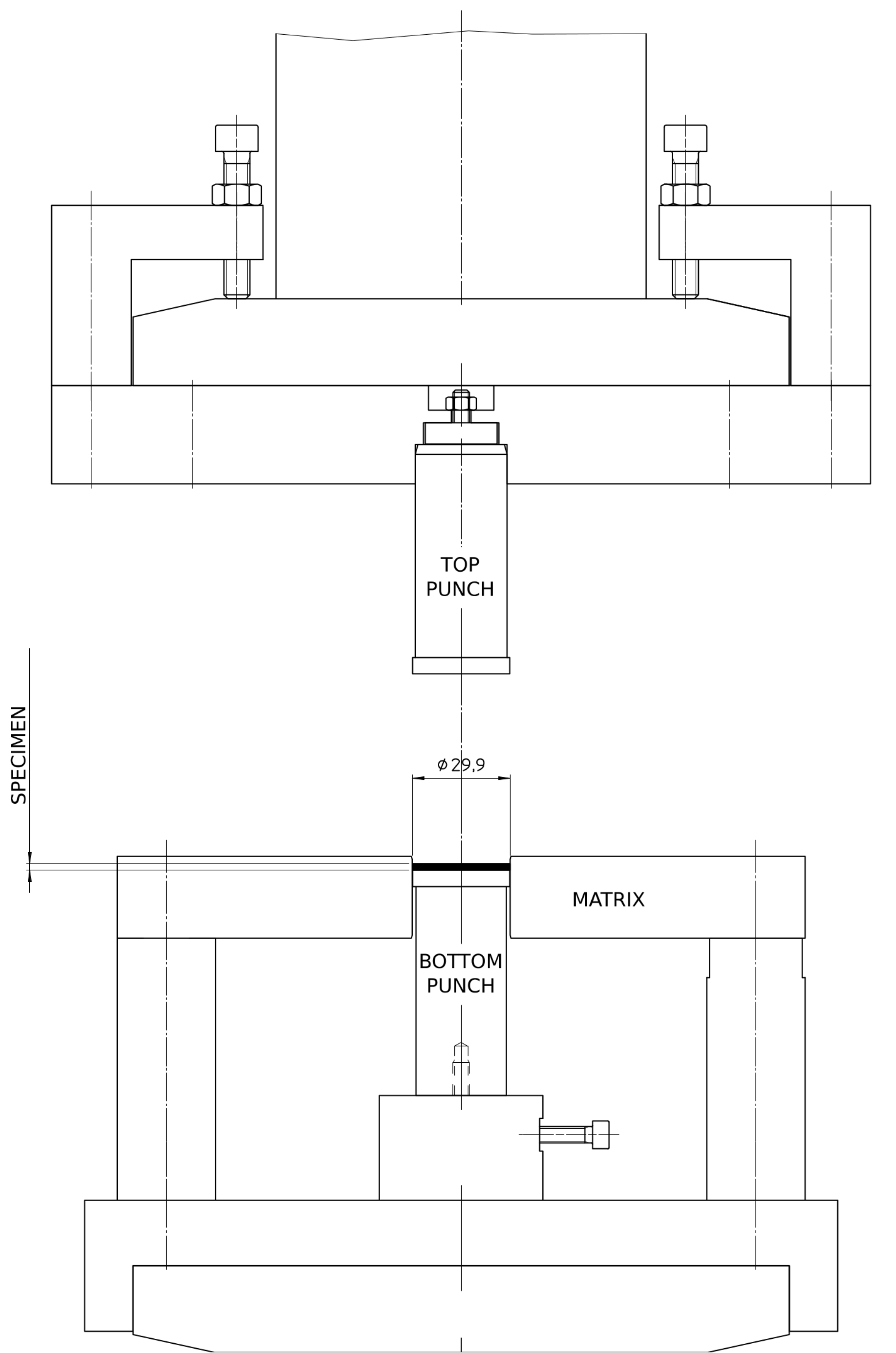}
\endminipage\hfil
\minipage{0.28\textwidth}
  \includegraphics[width=\linewidth]{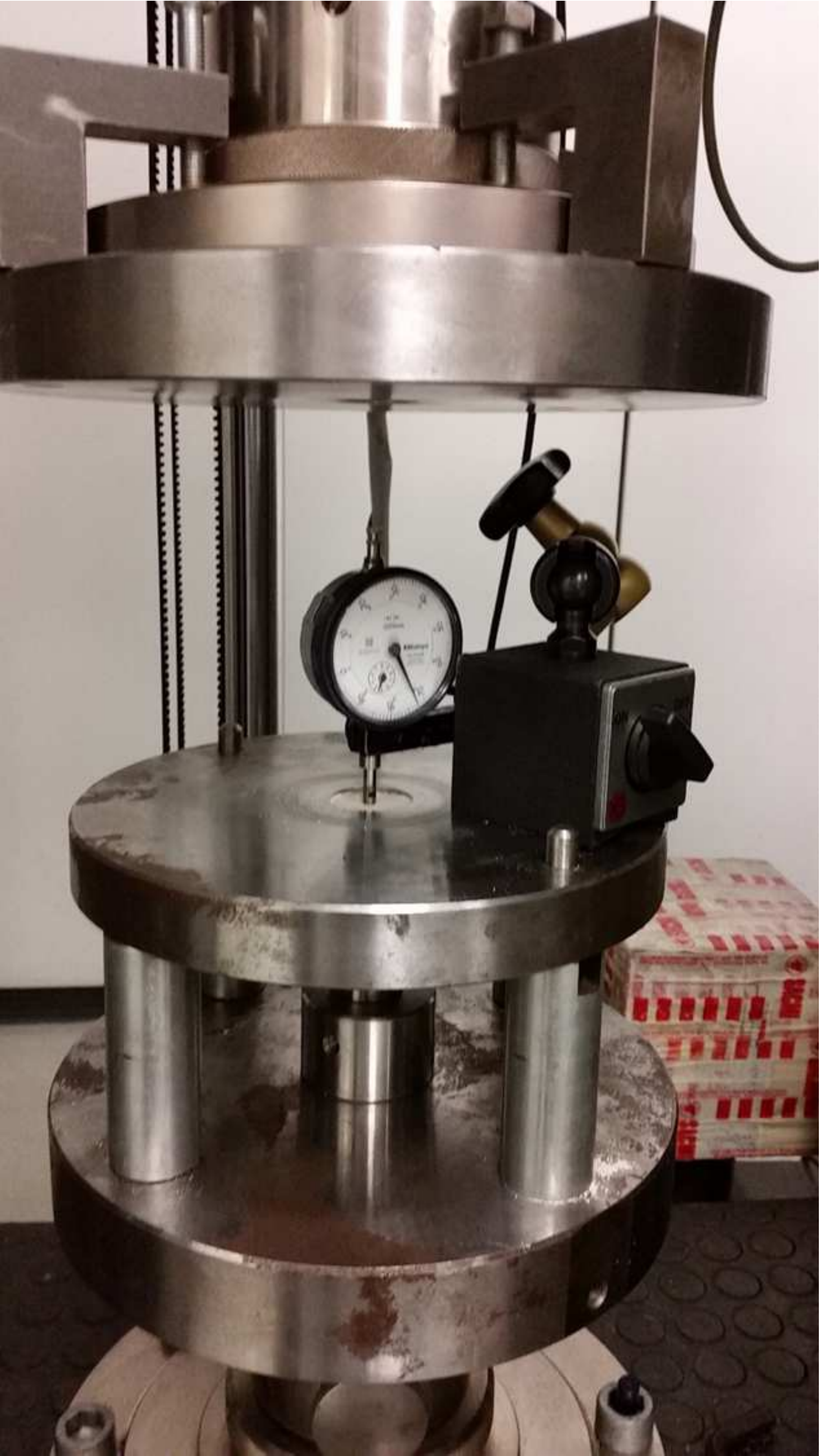}
\endminipage\hfil
\minipage{0.30\textwidth}
  \includegraphics[width=\linewidth]{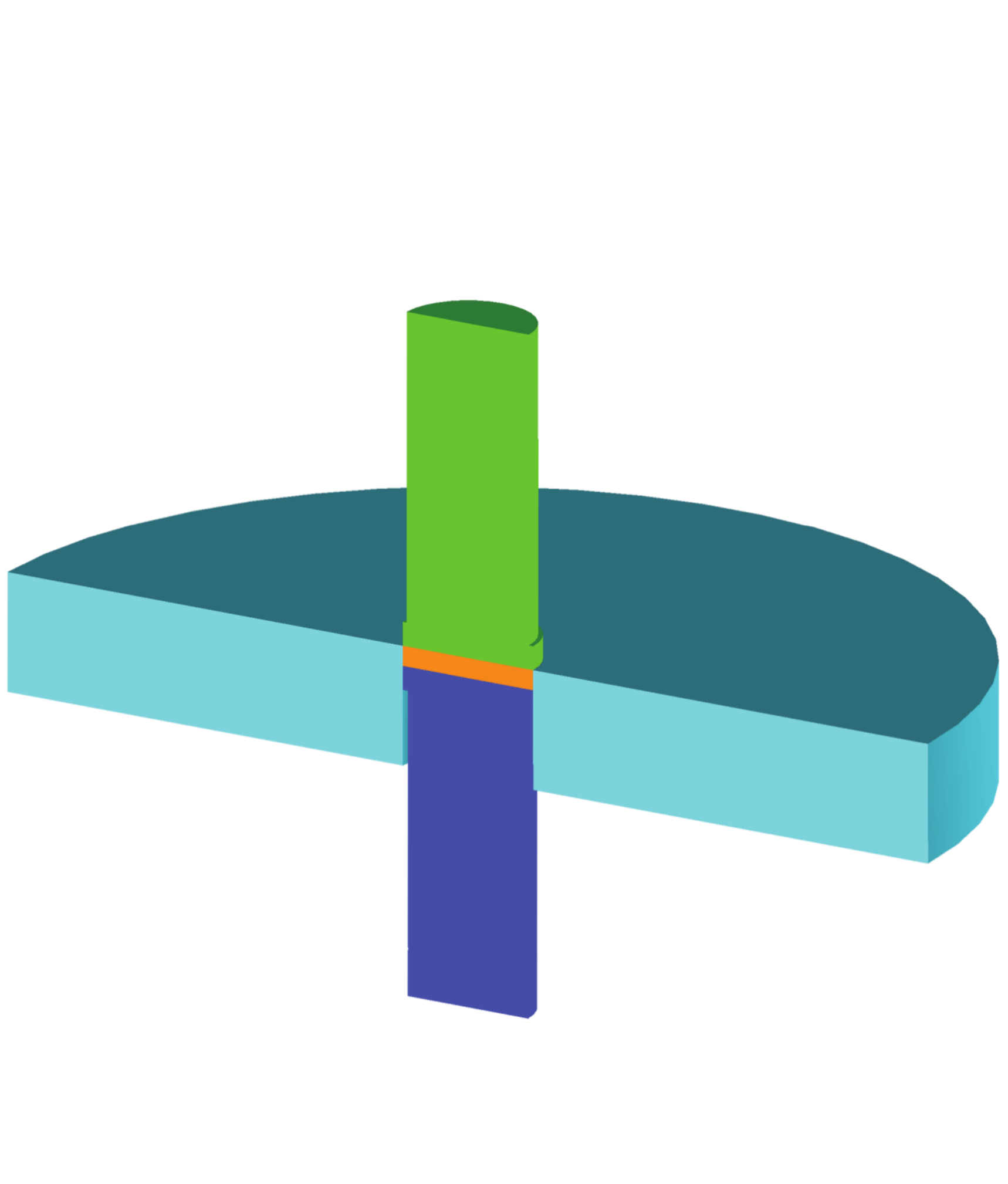}
\endminipage
\caption{\footnotesize Experimental set-up used in the uniaxial compaction test of I14730 aluminum silicate powder, performed by \citet{Bosi2013}. Cross section (left) and photograph (centre) of the forming device. The elements considered in the simulation: matrix, upper and lower punches, and ceramic powder (right).}
\label{expsetup}
\end{center}
\end{figure}

The experimental set-up used in the uniaxial compaction test of I14730 aluminum silicate powder, performed by \citet{Bosi2013} is shown in Fig.~\ref{expsetup}: the schematic cross section on the left and a photograph in the centre. The elements considered in the simulation are shown on the right: ceramic powder, cylindrical matrix, bottom and upper punches.

The simulation comprises the following steps: geostatic step, in which an initial confinement $p_0$ is imposed on the powder; uniaxial compaction phase, in which the upper punch is loaded by the given vertical pressure; unloading step, in which the load on the upper punch is removed; extraction, in which the matrix is removed in order to simulate the removal of the tablet from the cylindrical mould.

\begin{figure}[!htcb]
\begin{center}
\minipage{0.45\textwidth}
  \includegraphics[width=\linewidth]{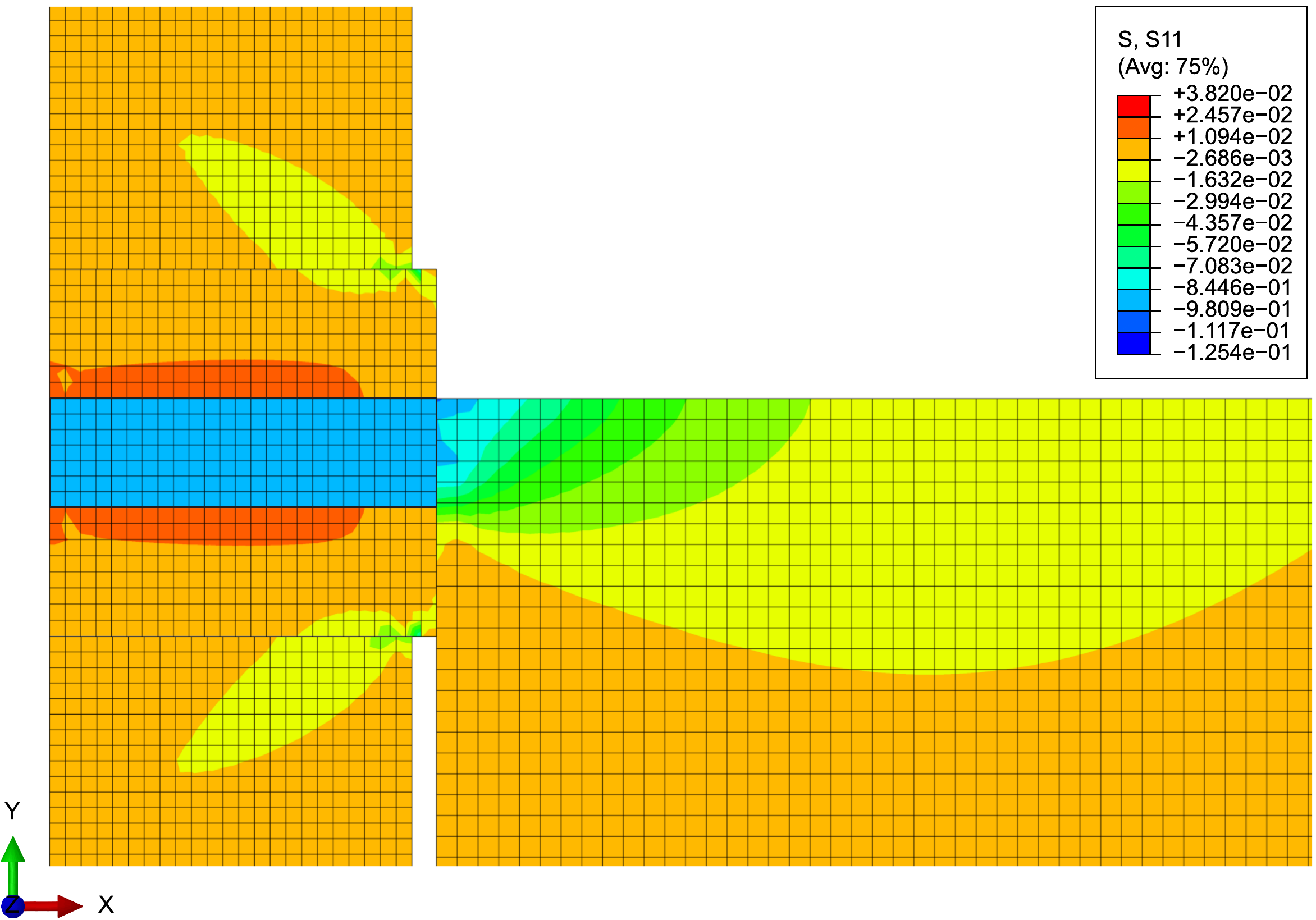}
    \caption*{\footnotesize \textbf{(a)} Geostatic step: $\sigma_{xx}$}
\endminipage\hfil
\minipage{0.45\textwidth}
  \includegraphics[width=\linewidth]{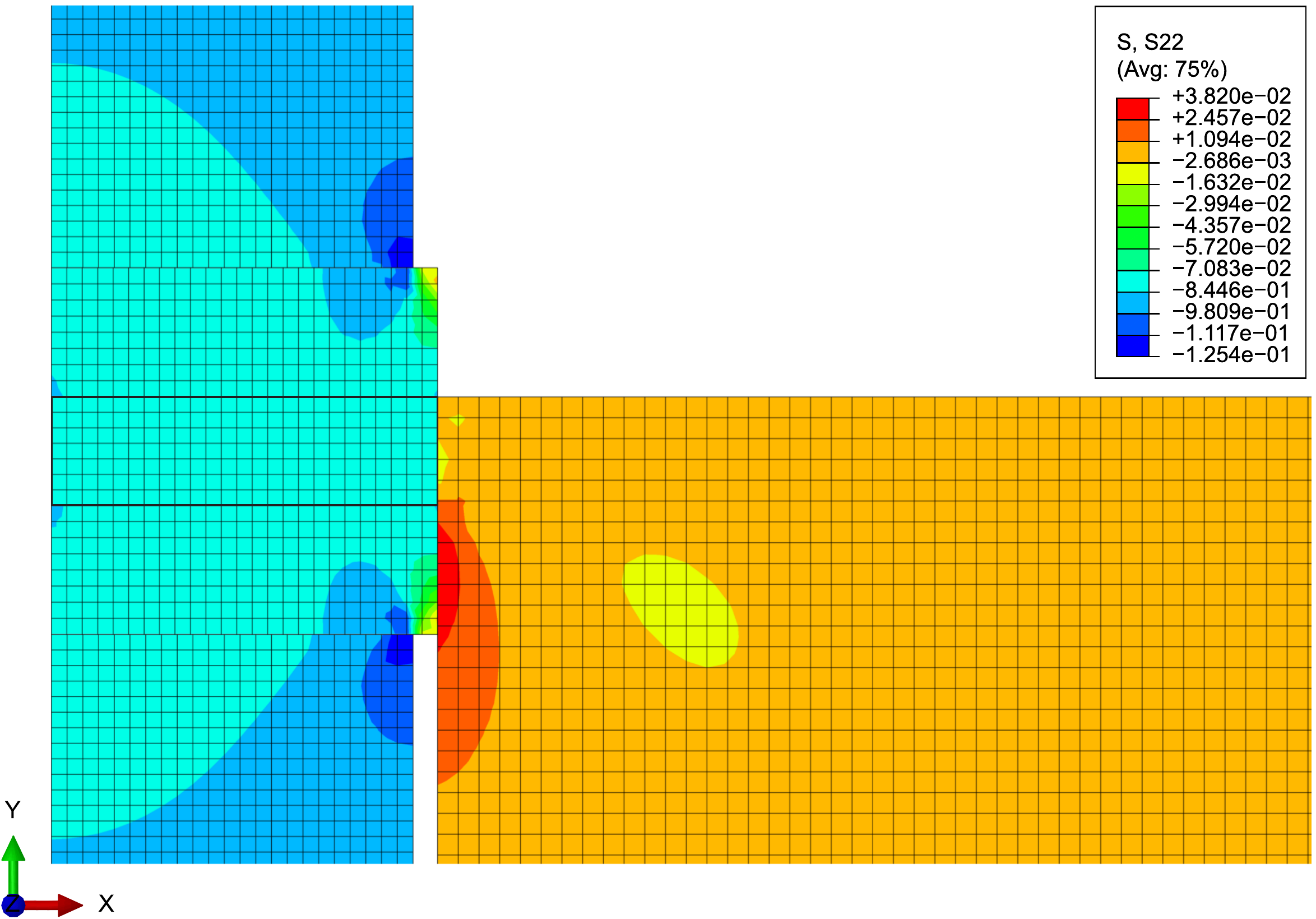}
    \caption*{\footnotesize \textbf{(b)} Geostatic step: $\sigma_{yy}$}
\endminipage\\[3mm]
\minipage{0.45\textwidth}
  \includegraphics[width=\linewidth]{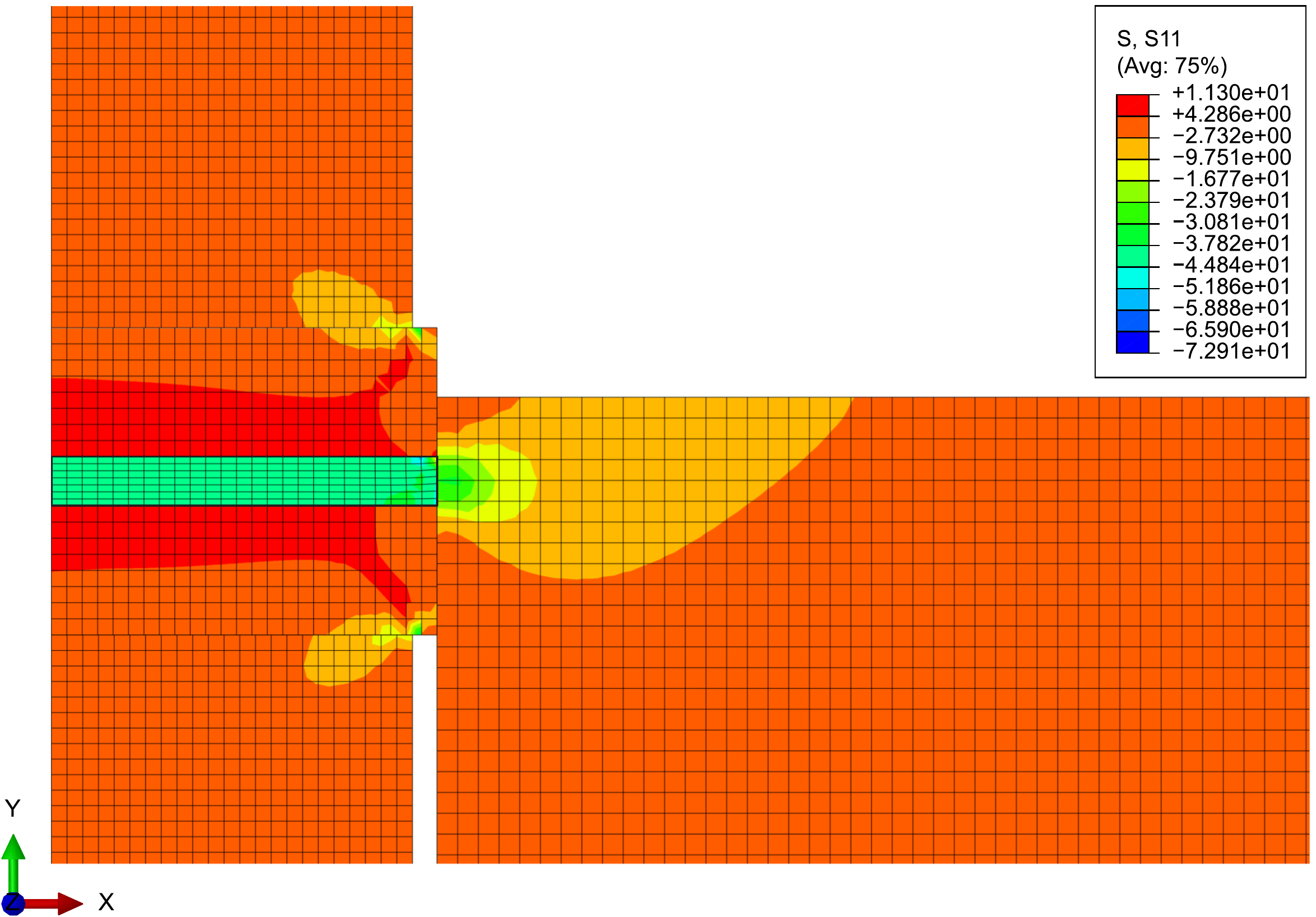}
    \caption*{\footnotesize \textbf{(c)} End of compaction phase: $\sigma_{xx}$}
\endminipage\hfil
\minipage{0.45\textwidth}
  \includegraphics[width=\linewidth]{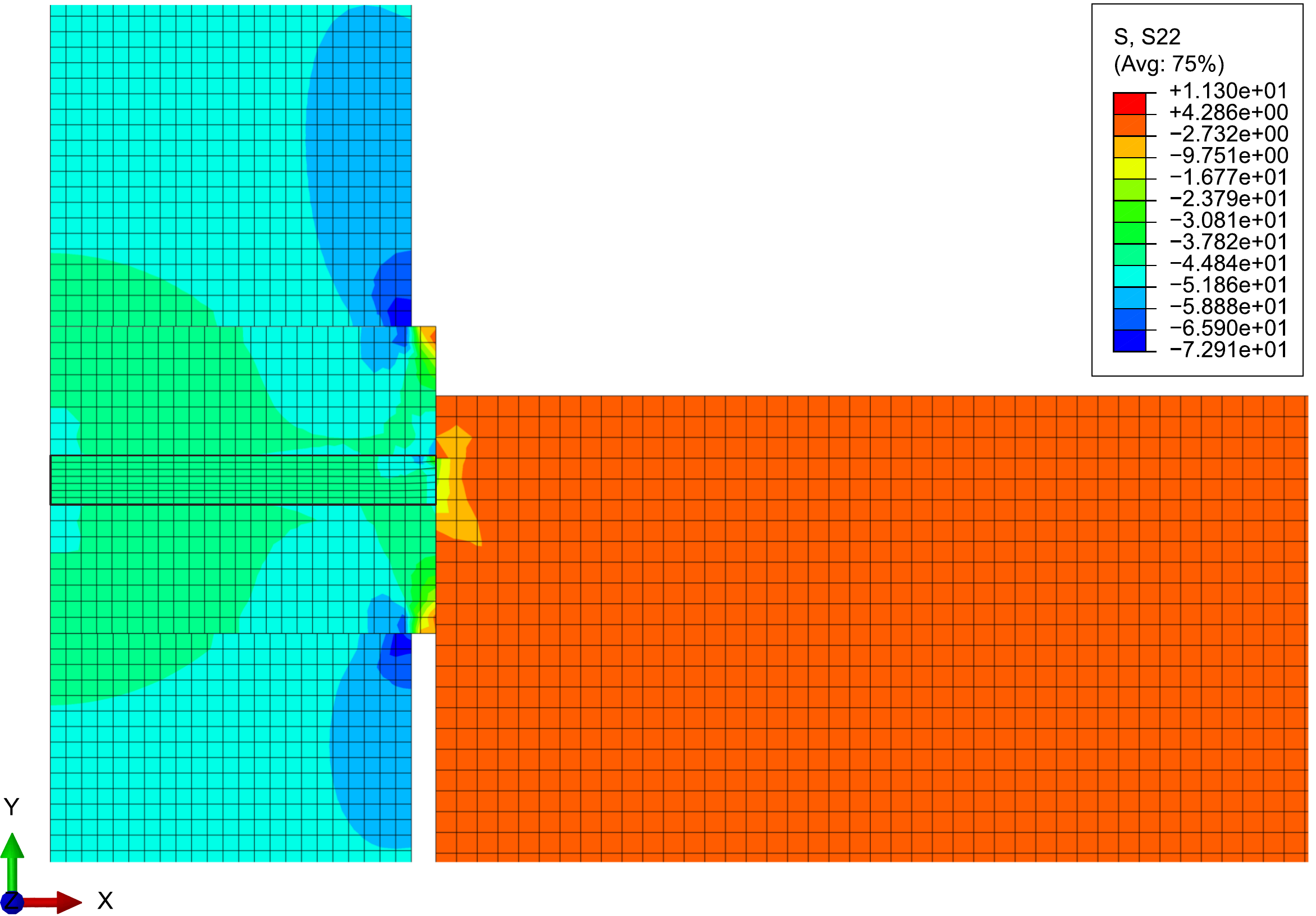}
\caption*{\footnotesize \textbf{(d)} End of compaction phase: $\sigma_{yy}$}
\endminipage\\[3mm]
\minipage{0.45\textwidth}
  \includegraphics[width=\linewidth]{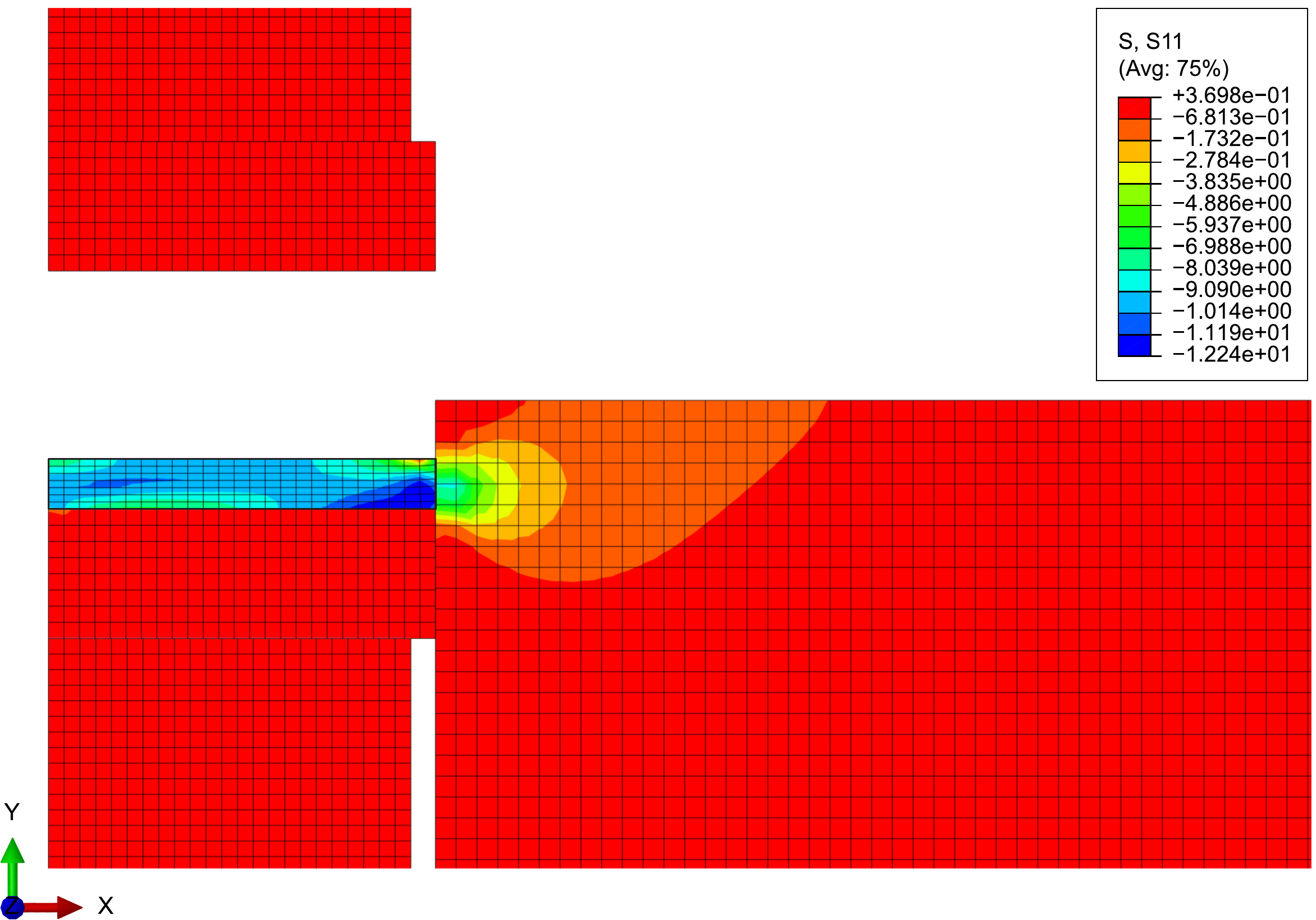}
\caption*{\footnotesize \textbf{(e)} Unloading: $\sigma_{xx}$}
\endminipage\hfil
\minipage{0.45\textwidth}
  \includegraphics[width=\linewidth]{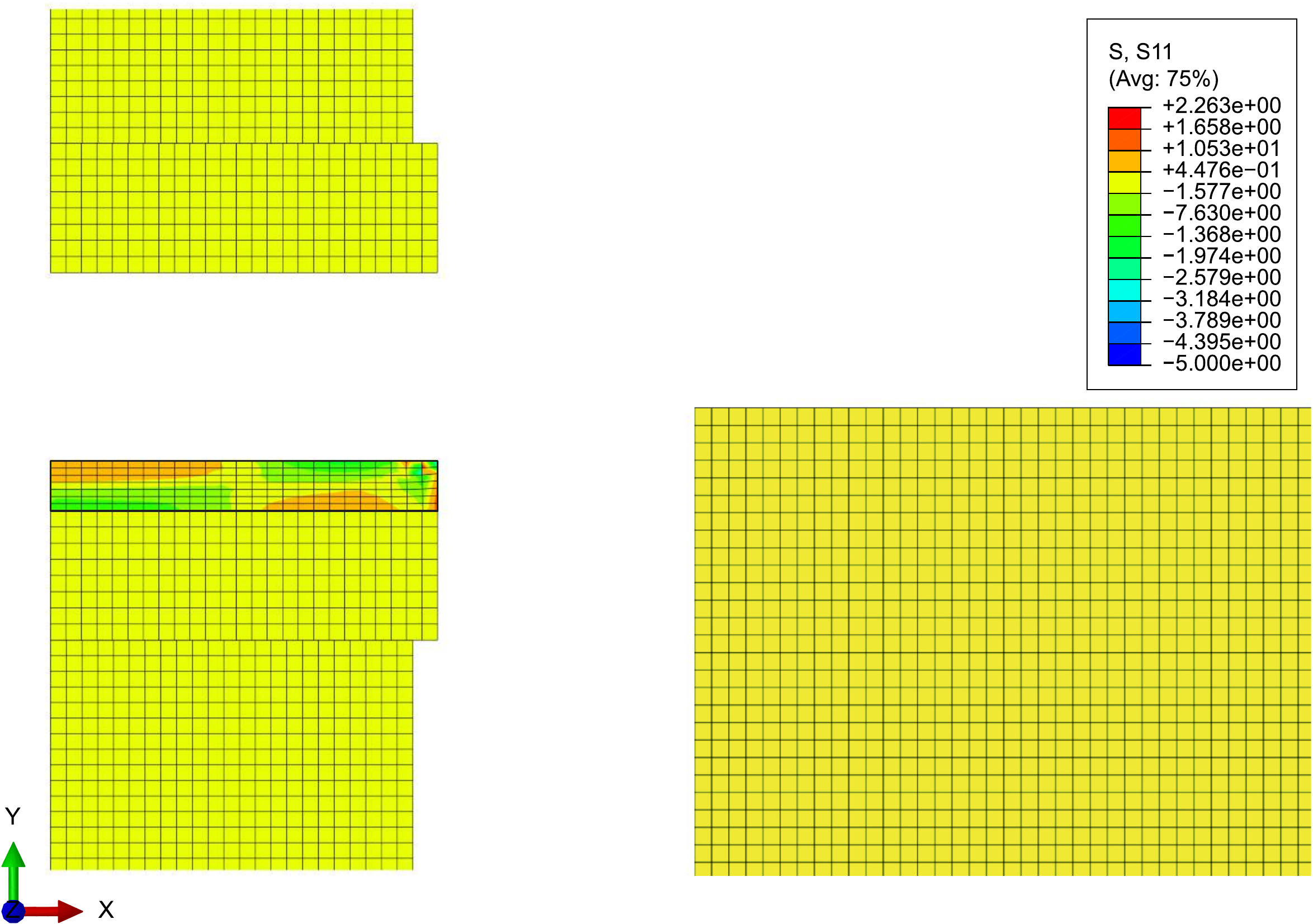}
\caption*{\footnotesize \textbf{(f)} Extraction of the green body: $\sigma_{xx}$}
\endminipage
\caption{\footnotesize Axisymmetric numerical simulation of tablet forming. Stress distribution (lateral stress $\sigma_{xx}$ and axial stress $\sigma_{yy}$) in the ceramic powder and in the mould (composed by matrix, upper and lower punches) at the end of the main stages of powder compaction: geostatic step (upper part), axial loading (central part), axial unloading (lower part, left) and extraction of the tablet (lower part, right).}
\label{2D-simulation}
\end{center}
\end{figure}

The stresses developing in the ceramic powder as well as in the mould are shown in Fig.~\ref{2D-simulation}. The analysis made also possible to investigate the influence of friction in the forming process and to determine the transversal pressure on the steel matrix.

The dimensions and densities of the formed tablets are reported in Tables \ref{densities-table-55} and \ref{densities-table-75}, for aluminum silicate powder with water content 5.5\% and 7.5\%, respectively. The simulation results are in good agreement with the experimental values. We note that a reduced radial ‘springback’ after extraction of the tablet from the mould was predicted by the numerical simulations.

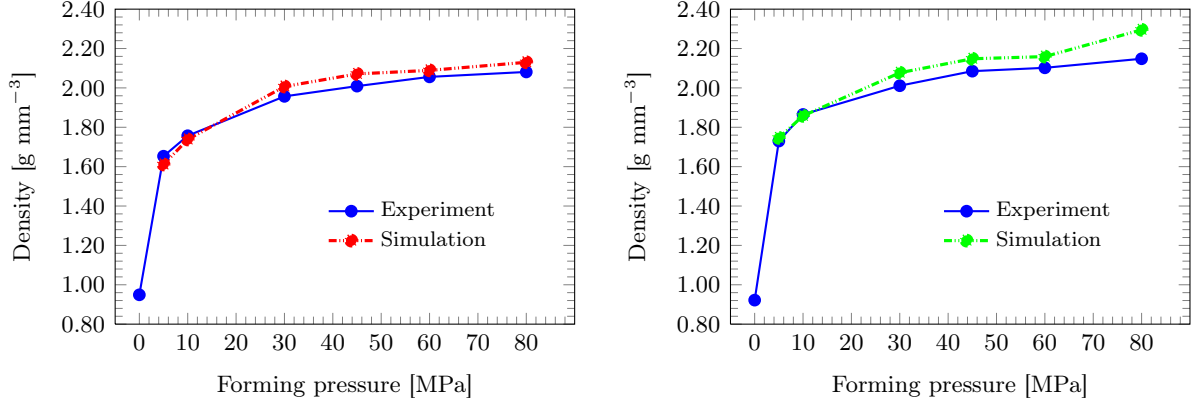
\begin{figure}[!htcb]
\centering
\begin{tikzpicture}
\begin{axis}
[scale only axis,                       
font=\footnotesize,                     
width=0.37\columnwidth,                 
height=0.20\textheight,                 
xmin=-5, xmax=90, ymin=0.8, ymax=2.4,     
scaled ticks=false,                     
  xticklabel style={
    /pgf/number format/precision=0,     
    /pgf/number format/fixed,
    /pgf/number format/fixed zerofill,  
  },
	yticklabel style={
    /pgf/number format/precision=2,     
    /pgf/number format/fixed,
    /pgf/number format/fixed zerofill,  
  },
    xtick={0,10,20,30,40,50,60,70,80}, 
    ytick={0.8,1.0,1.2,1.4,1.6,1.8,2.0,2.2,2.4},  
    minor x tick num = 4,               
    minor y tick num = 4,               
    xlabel={Forming pressure [\si{\mega Pa}]},
    ylabel={Density [\si{g \milli m^{-3}}] }, 
  legend style={
  at={(0.65,0.2)},
  anchor=south,
  legend columns=1,
  cells={anchor=west},
  font=\scriptsize,
	draw=none,
}]

\addplot [blue,thick,mark=*]
file {density-experiment-55-2.txt};
\addplot [red,densely dashdotdotted,very thick,mark=*]
file {density-simulations-55-2.txt};
\legend{Experiment, Simulation}
\end{axis}
\end{tikzpicture}
\hfil
\begin{tikzpicture}
\begin{axis}
[scale only axis,                       
font=\footnotesize,                     
width=0.37\columnwidth,                 
height=0.20\textheight,                 
xmin=-5, xmax=90, ymin=0.8, ymax=2.4,     
scaled ticks=false,                     
  xticklabel style={
    /pgf/number format/precision=0,     
    /pgf/number format/fixed,
    /pgf/number format/fixed zerofill,  
  },
	yticklabel style={
    /pgf/number format/precision=2,     
    /pgf/number format/fixed,
    /pgf/number format/fixed zerofill,  
  },
    xtick={0,10,20,30,40,50,60,70,80}, 
    ytick={0.8,1.0,1.2,1.4,1.6,1.8,2.0,2.2,2.4},  
    minor x tick num = 4,               
    minor y tick num = 4,               
    xlabel={Forming pressure [\si{\mega Pa}]},
    ylabel={Density [\si{g \milli m^{-3}}] }, 
  legend style={
  at={(0.65,0.2)},
  anchor=south,
  legend columns=1,
  cells={anchor=west},
  font=\scriptsize,
	draw=none,
}]

\addplot [blue,thick,mark=*]
file {density-experiment-75.txt};
\addplot [green,densely dashdotdotted,very thick,mark=*]
file {density-simulations-75.txt};
\legend{Experiment, Simulation}
\end{axis}
\end{tikzpicture}
\caption[]{\footnotesize Densities of I14730 aluminium silicate tablets, with 5.5\% and 7.5\% of water content: comparison between experimental result and numerical simulation for different forming pressures. The first experimental point shows the average bulk density of the ceramic powder when inserted in the cylindrical mould.}
\label{densities-comparison}
\end{figure}

A comparison of experimental and simulated compact densities is shown in Fig.~\ref{densities-comparison}, for water content 5.5\% (left) and 7.5\% (right).

\clearpage

%
\begin{table}[tp]
\centering
\caption[Densities after extraction]{\footnotesize Dimensions and densities of I14730 aluminum silicate tablets (w=5.5\%) after completion of the uniaxial compaction process and extraction from the cylindrical mould.}
\label{densities-table-55}
\begin{tabular}{cccccc}
\toprule
  \parbox{25mm}{\centering Forming \\ pressure \normalsize{[MPa]}} 
&
& \parbox{20mm}{\centering Area \\ \normalsize{[mm$^2$]}} 
& \parbox{20mm}{\centering Height \\ \normalsize{[mm]}} 
& \parbox{20mm}{\centering Diameter \\ \normalsize{[mm]}} 
& \parbox{20mm}{\centering Density \\ \normalsize{[g cm$^{-3}$]}} \\
\midrule
  \multirow{2}*{\text{5}}	& Experiment	& \num{706.387}	& \num{2.338}	& \num{29.990}	& \num{1.653} \\
				& Simulation	& \num{708.452}	& \num{2.392}	& \num{30.034}	& \num{1.611} \\
\midrule
  \multirow{2}*{\text{10}}	& Experiment	& \num{706.858}	& \num{2.198}	& \num{30.000}	& \num{1.757} \\
				& Simulation	& \num{705.841}	& \num{2.225}	& \num{29.978}	& \num{1.738} \\
\midrule
  \multirow{2}*{\text{30}}	& Experiment	& \num{708.273}	& \num{2.042}	& \num{30.030}	& \num{1.957} \\
				& Simulation	& \num{703.554}	& \num{2.004}	& \num{29.930}	& \num{2.007} \\
\midrule
  \multirow{2}*{\text{45}}	& Experiment	& \num{708.745}	& \num{1.966}	& \num{30.040}	& \num{2.009} \\
				& Simulation	& \num{703.225}	& \num{1.923}	& \num{29.923}	& \num{2.071} \\
\midrule
  \multirow{2}*{\text{60}}	& Experiment	& \num{708.745}	& \num{1.880}	& \num{30.040}	& \num{2.056} \\
				& Simulation	& \num{703.065}	& \num{1.865}	& \num{29.919}	& \num{2.089} \\
\midrule
  \multirow{2}*{\text{80}}	& Experiment	& \num{708.745}	& \num{1.844}	& \num{30.040}	& \num{2.081} \\
				& Simulation	& \num{702.962}	& \num{1.817}	& \num{29.917}	& \num{2.130} \\
\bottomrule
\end{tabular}
\end{table}
%
\begin{table}[tp]
\centering
\caption[Densities after extraction]{\footnotesize Dimensions and densities of I14730 aluminum silicate tablets (w=7.5\%) after completion of the uniaxial compaction process and extraction from the cylindrical mould.}
\label{densities-table-75}
\begin{tabular}{cccccc}
\toprule
\parbox{25mm}{\centering Forming \\ pressure \normalsize{[MPa]}} 
&
& \parbox{20mm}{\centering Area \\ \normalsize{[mm$^2$]}} 
& \parbox{20mm}{\centering Height \\ \normalsize{[mm]}} 
& \parbox{20mm}{\centering Diameter \\ \normalsize{[mm]}} 
& \parbox{20mm}{\centering Density \\ \normalsize{[g cm$^{-3}$]}} \\
\midrule
  \multirow{2}*{\text{5}}	& Experiment	& \num{705.916}	& \num{2.236}	& \num{29.980}	& \num{1.730} \\
				& Simulation	& \num{704.175}	& \num{2.221}	& \num{29.943}	& \num{1.746} \\
\midrule
  \multirow{2}*{\text{10}}	& Experiment	& \num{705.916}	& \num{2.074}	& \num{29.980}	& \num{1.865} \\
				& Simulation	& \num{703.451}	& \num{2.089}	& \num{29.928}	& \num{1.858} \\
\midrule
  \multirow{2}*{\text{30}}	& Experiment	& \num{707.801}	& \num{1.918}	& \num{30.020}	& \num{2.011} \\
				& Simulation	& \num{702.981}	& \num{1.870}	& \num{29.918}	& \num{2.077} \\
\midrule
  \multirow{2}*{\text{45}}	& Experiment	& \num{707.943}	& \num{1.836}	& \num{30.023}	& \num{2.085} \\
				& Simulation	& \num{702.905}	& \num{1.795}	& \num{29.916}	& \num{2.148} \\
\midrule
  \multirow{2}*{\text{60}}	& Experiment	& \num{708.745}	& \num{1.772}	& \num{30.040}	& \num{2.102} \\
				& Simulation	& \num{702.868}	& \num{1.740}	& \num{29.915}	& \num{2.159} \\
\midrule
  \multirow{2}*{\text{80}}	& Experiment	& \num{708.273}	& \num{1.814}	& \num{30.030}	& \num{2.148} \\
				& Simulation	& \num{702.868}	& \num{1.711}	& \num{29.915}	& \num{2.295} \\
\bottomrule
\end{tabular}
\end{table}

\clearpage

\subsection{Three-dimensional numerical simulation of an industrial tile forming process}
\label{sec06}

Three-dimensional FE simulations of an industrial tile forming process have been performed in Abaqus FEA environment using the developed constitutive model and the identified parameters listed in Tab.~\ref{tab:alluminium-silicate-parameters}. 
The 3D model involves contact interactions between ceramic powder, steel matrix and top/bottom steel plates, where the friction coefficient reported in Sec.~\ref{friction} has been used.

\begin{figure}[!htcb]
\begin{center}
\minipage{0.48\textwidth}
  \includegraphics[width=\linewidth]{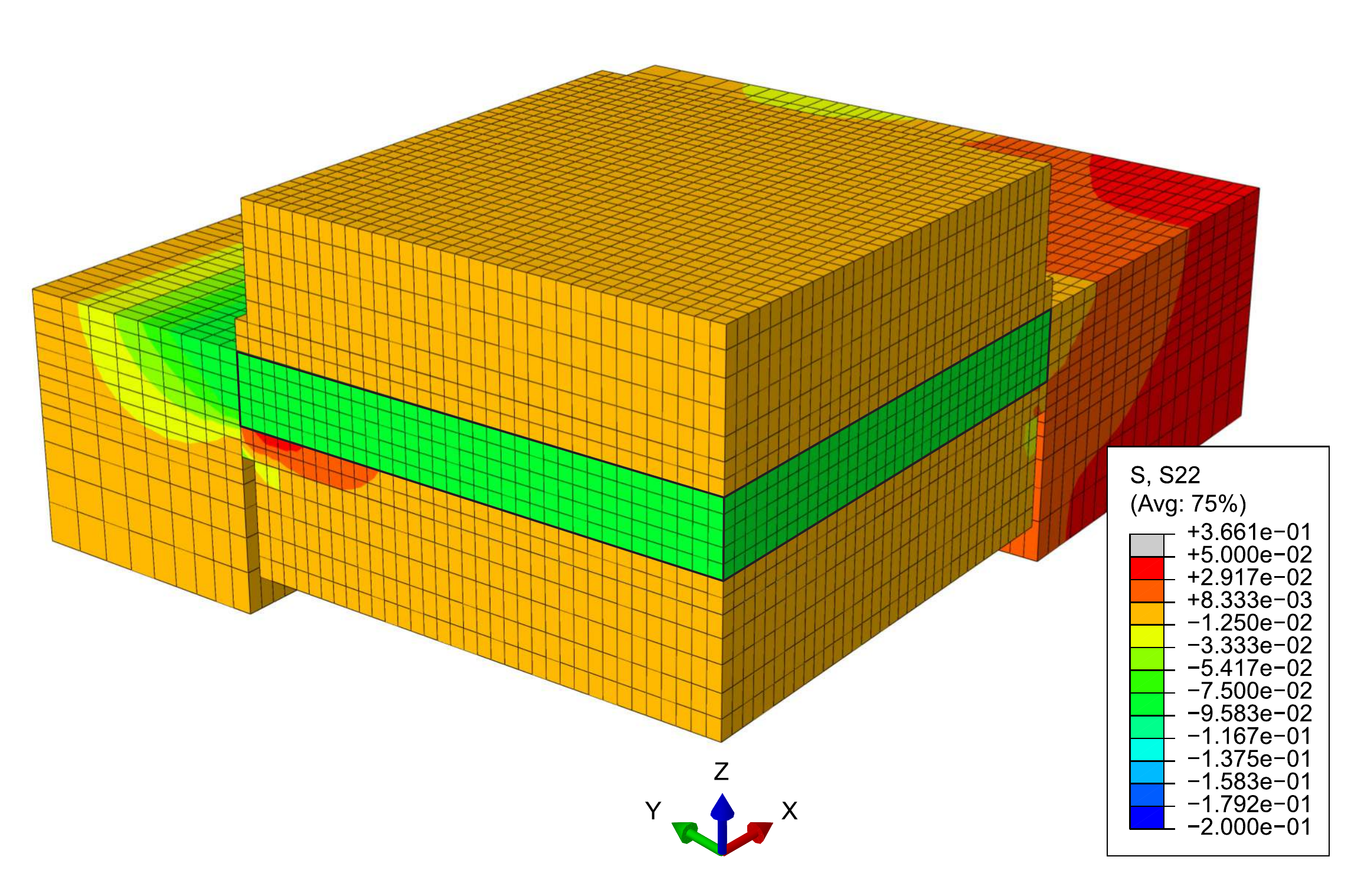}
  \caption*{\footnotesize \textbf{(a)} Geostatic step: $\sigma_{yy}$}
\endminipage\hfil
\minipage{0.48\textwidth}
  \includegraphics[width=\linewidth]{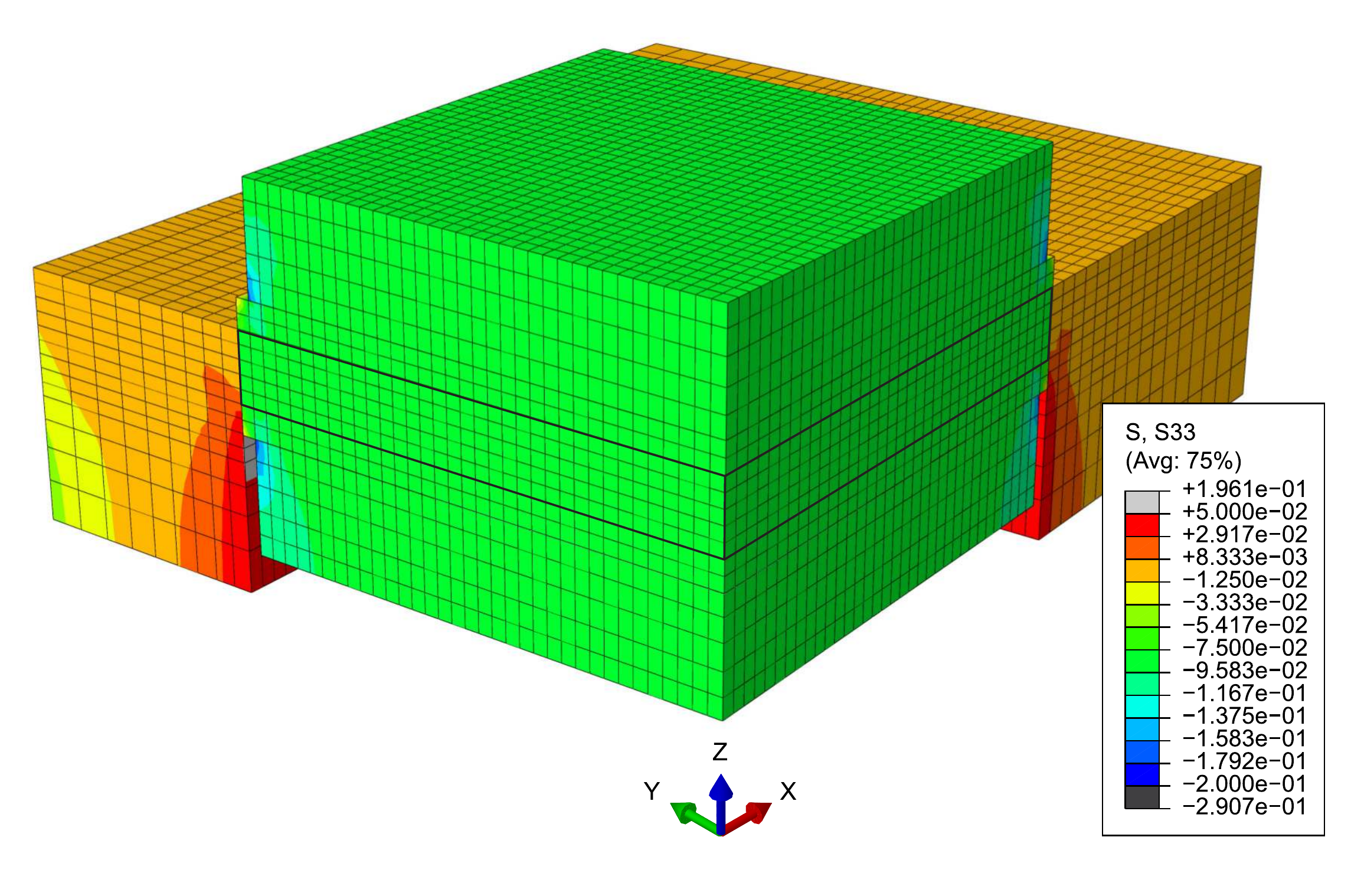}
  \caption*{\footnotesize \textbf{(b)} Geostatic step: $\sigma_{zz}$}
\endminipage \\[3mm]
\minipage{0.48\textwidth}
  \includegraphics[width=\linewidth]{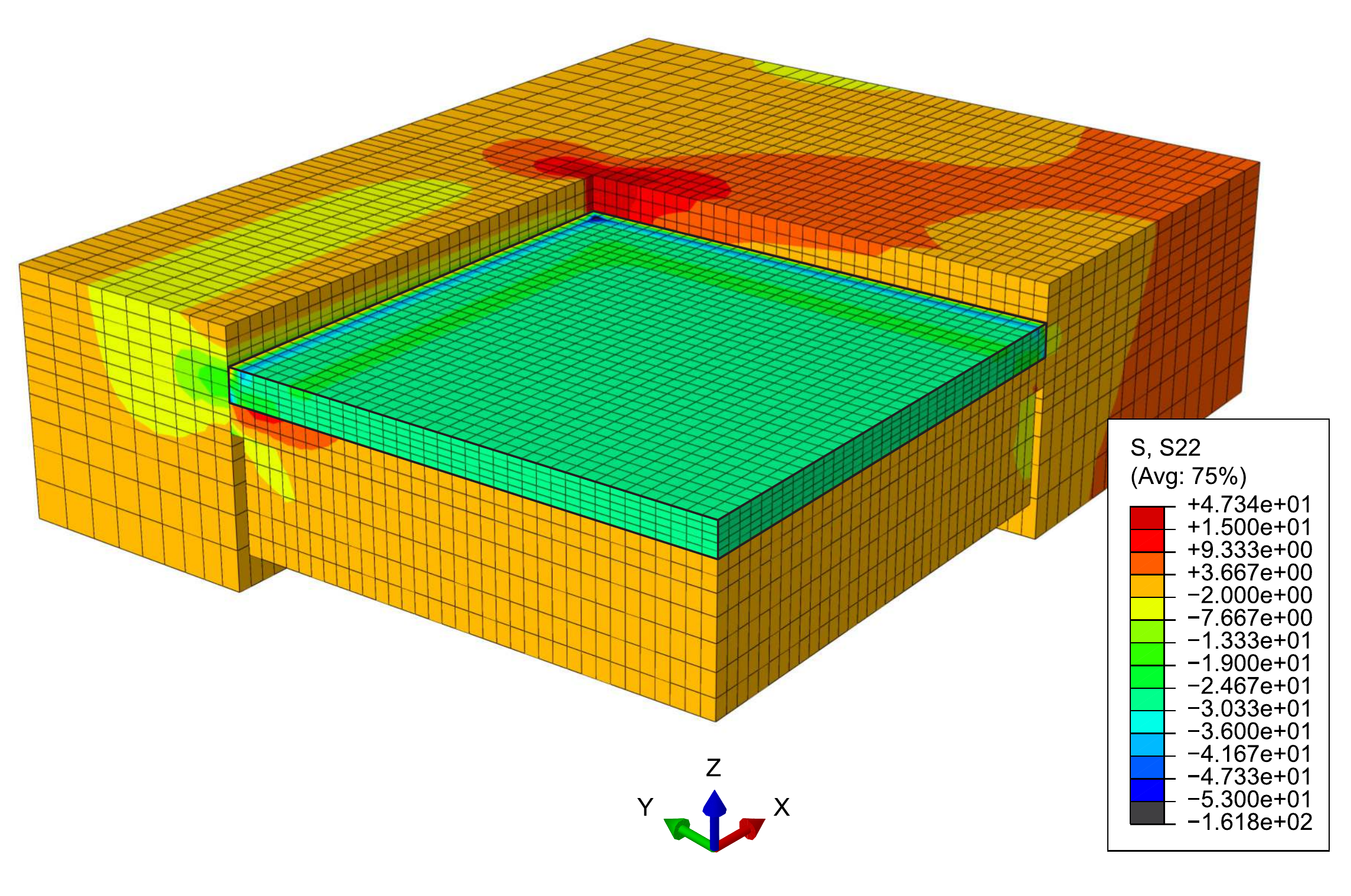}
  \caption*{\footnotesize \textbf{(c)} End of compaction phase: $\sigma_{yy}$}
\endminipage\hfil
\minipage{0.48\textwidth}
  \includegraphics[width=\linewidth]{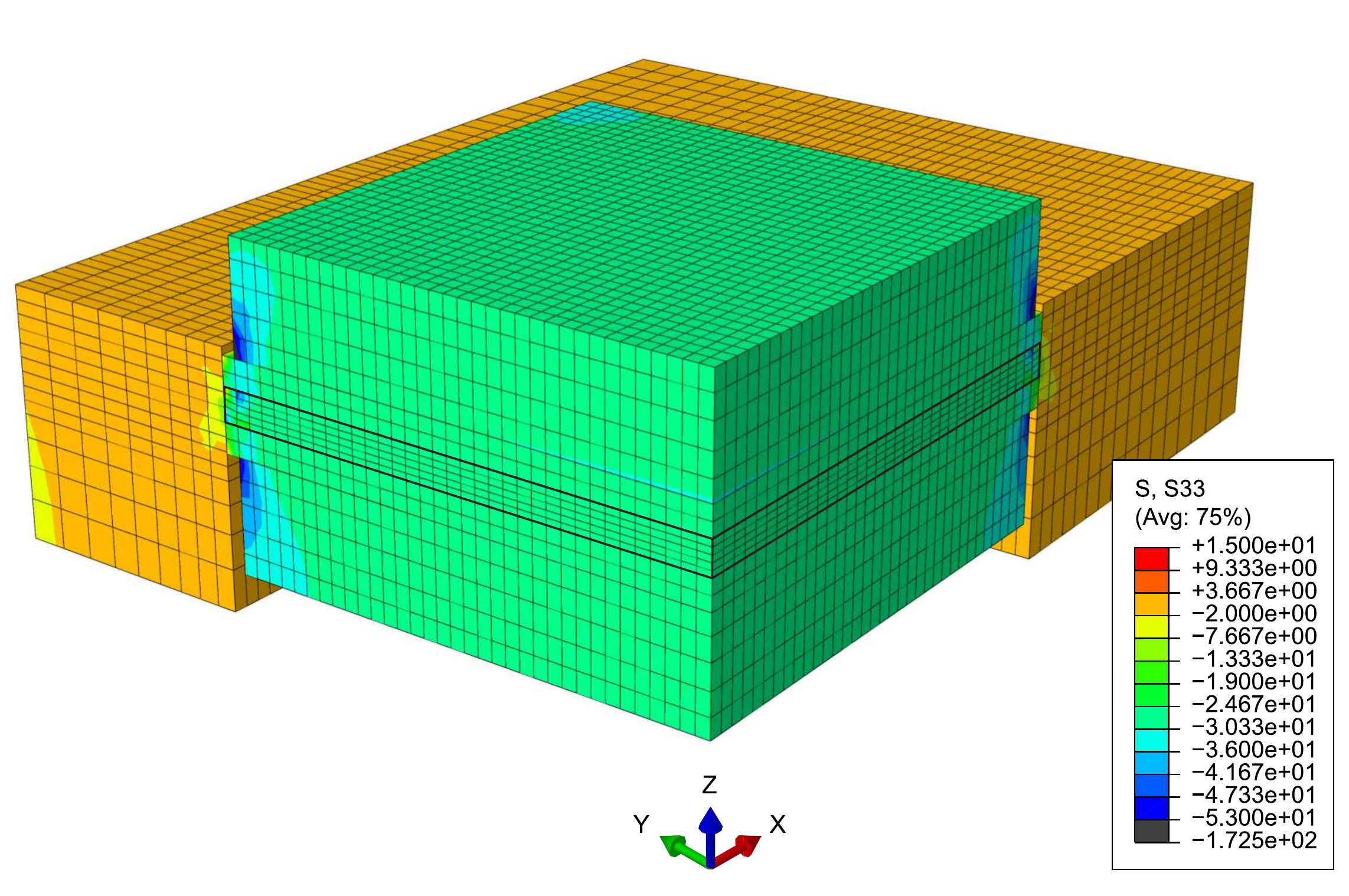}
  \caption*{\footnotesize \textbf{(d)} End of compaction phase: $\sigma_{zz}$}
\endminipage \\[3mm]
\minipage{0.48\textwidth}
  \includegraphics[width=\linewidth]{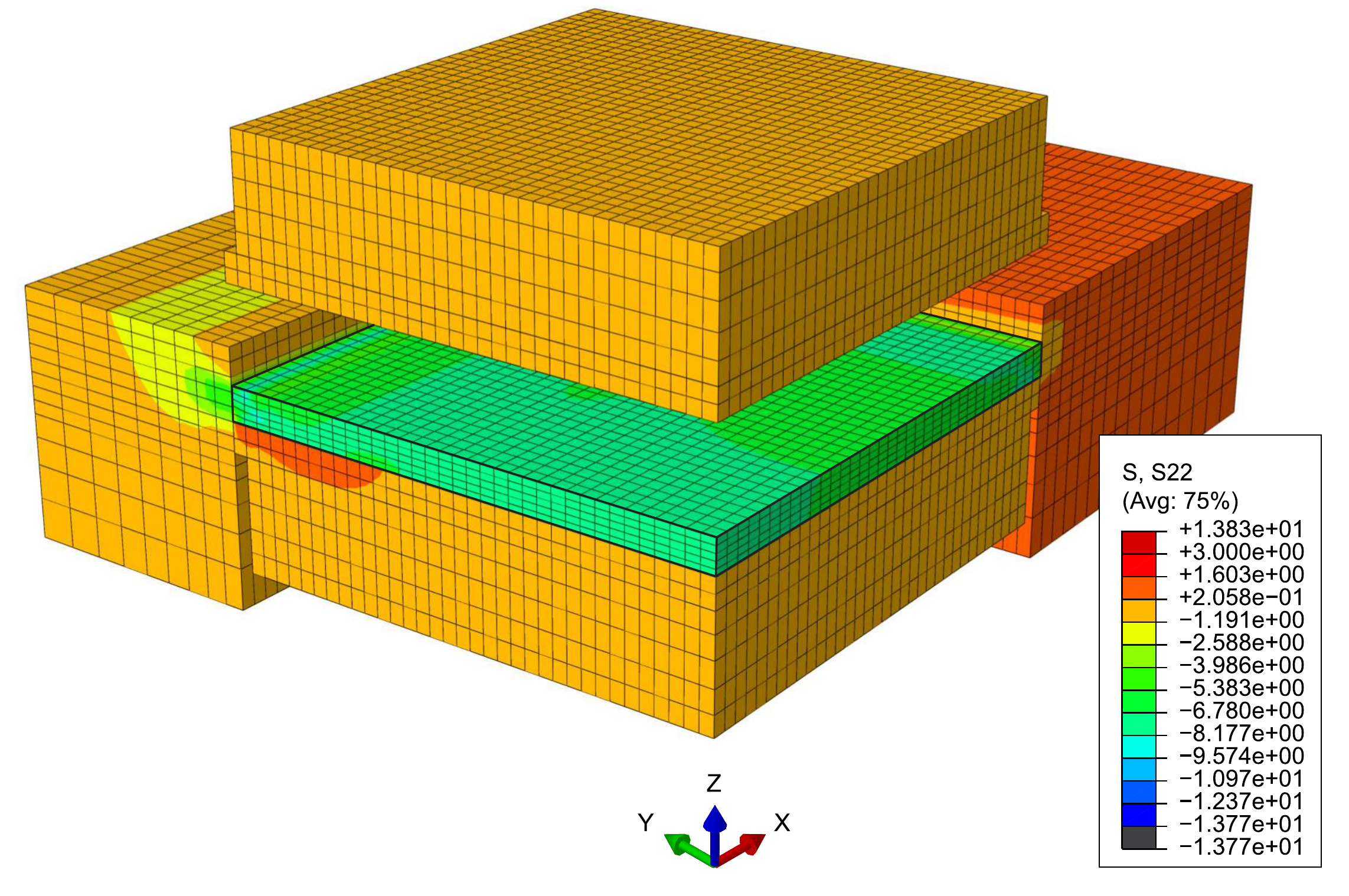}
  \caption*{\footnotesize \textbf{(e)} Unloading: $\sigma_{yy}$}
\endminipage\hfil
\minipage{0.48\textwidth}
  \includegraphics[width=\linewidth]{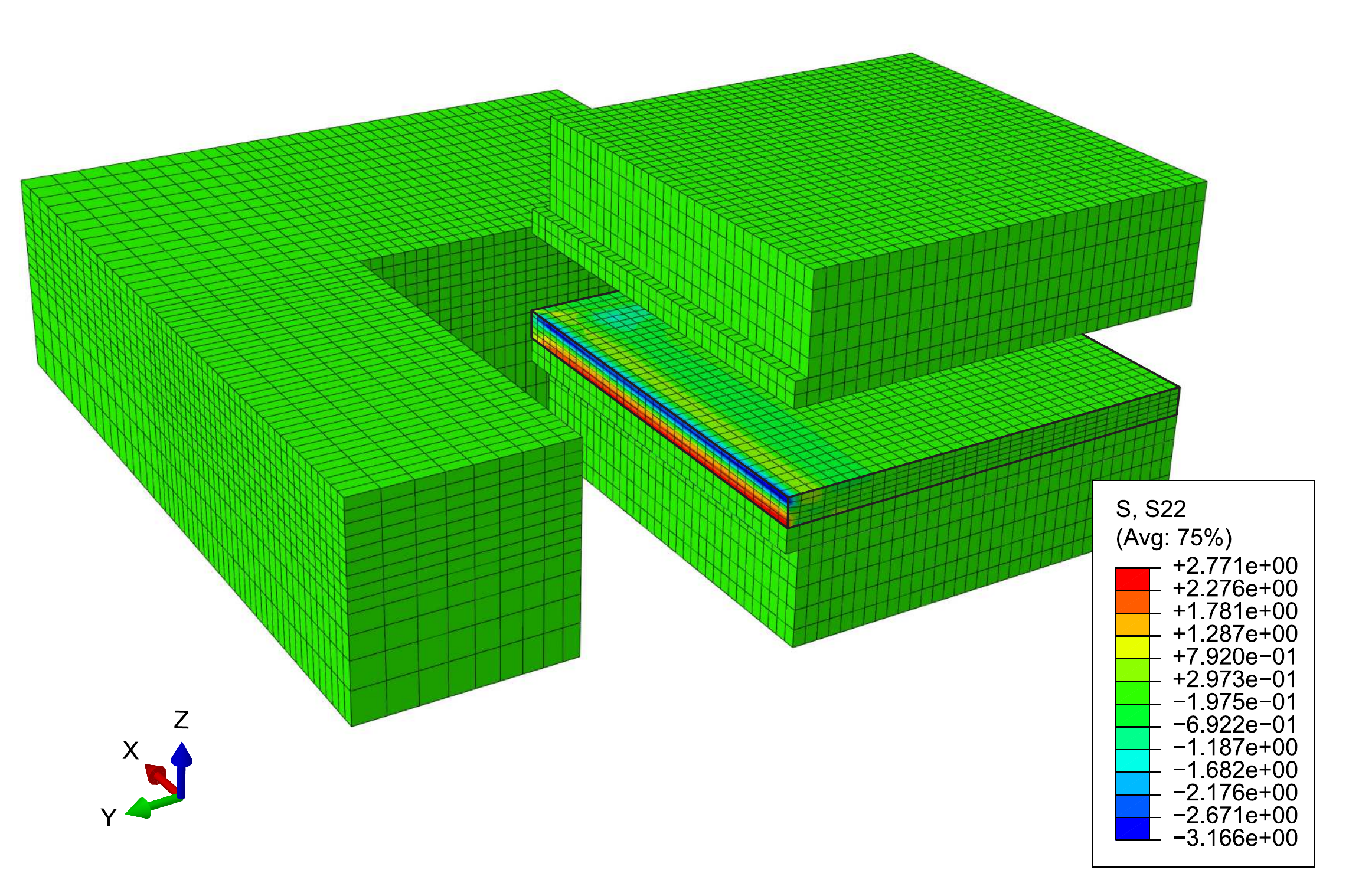}
  \caption*{\footnotesize \textbf{(f)} Extraction of the green body: $\sigma_{yy}$}
\endminipage
\caption{\footnotesize Three-dimensional numerical simulation of industrial tile forming. Stress distribution (lateral stress $\sigma_{yy}$ and axial stress $\sigma_{zz}$) in the ceramic powder and in the mould (composed by matrix, upper and lower plates) at the end of the main stages of powder compaction: geostatic step (upper part), axial loading (central part), axial unloading (lower part, left) and extraction of the tile (lower part, right).}
\label{3dimensional}
\end{center}
\end{figure}

Figure~\ref{3dimensional} shows the vertical and transverse stress distributions in the ceramic powder and in the mould (composed by matrix, upper and lower plates) at the end of the main stages of powder compaction. 

The initial values of isotropic stress and void ratio prescribed in the geostatic step are $p_0=0.9$ MPa and $e_0=2.04$, respectively, which are the same values used in the axisymmetric case. This step corresponds to the initial confinement inside the mould, see Fig.~\ref{3dimensional}a and \ref{3dimensional}b. 

At the end of the compaction phase, Fig.~\ref{3dimensional}c and \ref{3dimensional}d, we notice a non-uniform stress distribution, especially for the transversal stress $\sigma_{yy}$, in the external part of the tile, which is in contact with the die wall. This effect is mainly due to friction between the ceramic powder and the mould.

After the unloading step (removal of the upper plate), see Fig.~\ref{3dimensional}e, the transversal stress $\sigma_{yy}$ in the green tile is still quite high, approximately 8 MPa, due to the lateral constraint given by the matrix.

The residual stresses at the end of the extraction phase, predicted by the finite element simulation, are shown in Fig.~\ref{3dimensional}f. We notice a compressive transversal stress $\sigma_{yy}$ at the upper edge of the tile, which is balanced by a tensile $\sigma_{yy}$ at the lower edge. The capabilities of the model to predict residual stresses are crucial, since residual stresses may lead to fracture of the green body, by end capping or lamination.

\clearpage

\subsection{Estimation of transversal load on the lateral die wall}

\begin{figure}[!htcb]
\minipage{0.49\textwidth}
  \includegraphics[width=\linewidth]{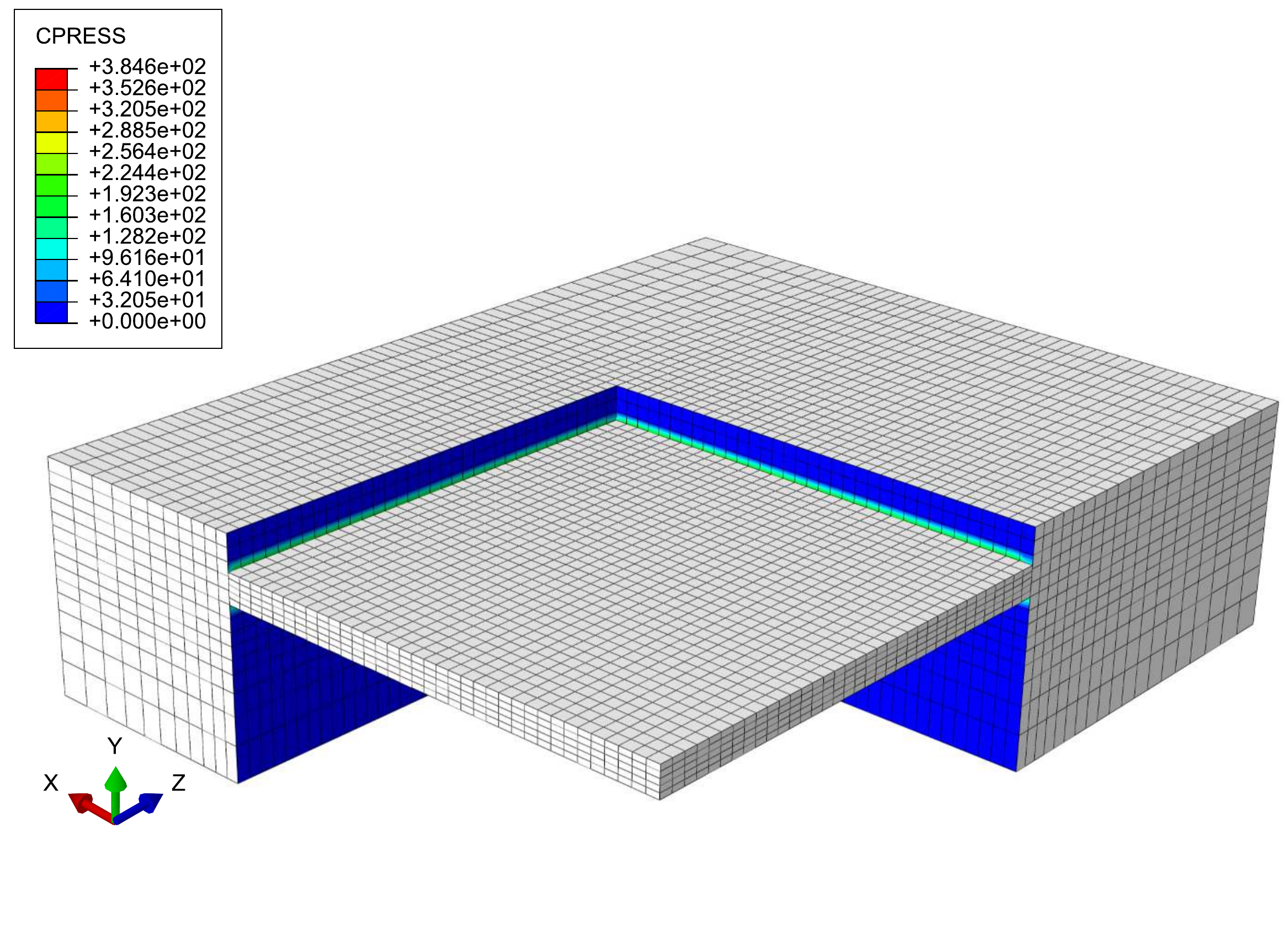}
\endminipage\hfil
\minipage{0.49\textwidth}
  \includegraphics[width=\linewidth]{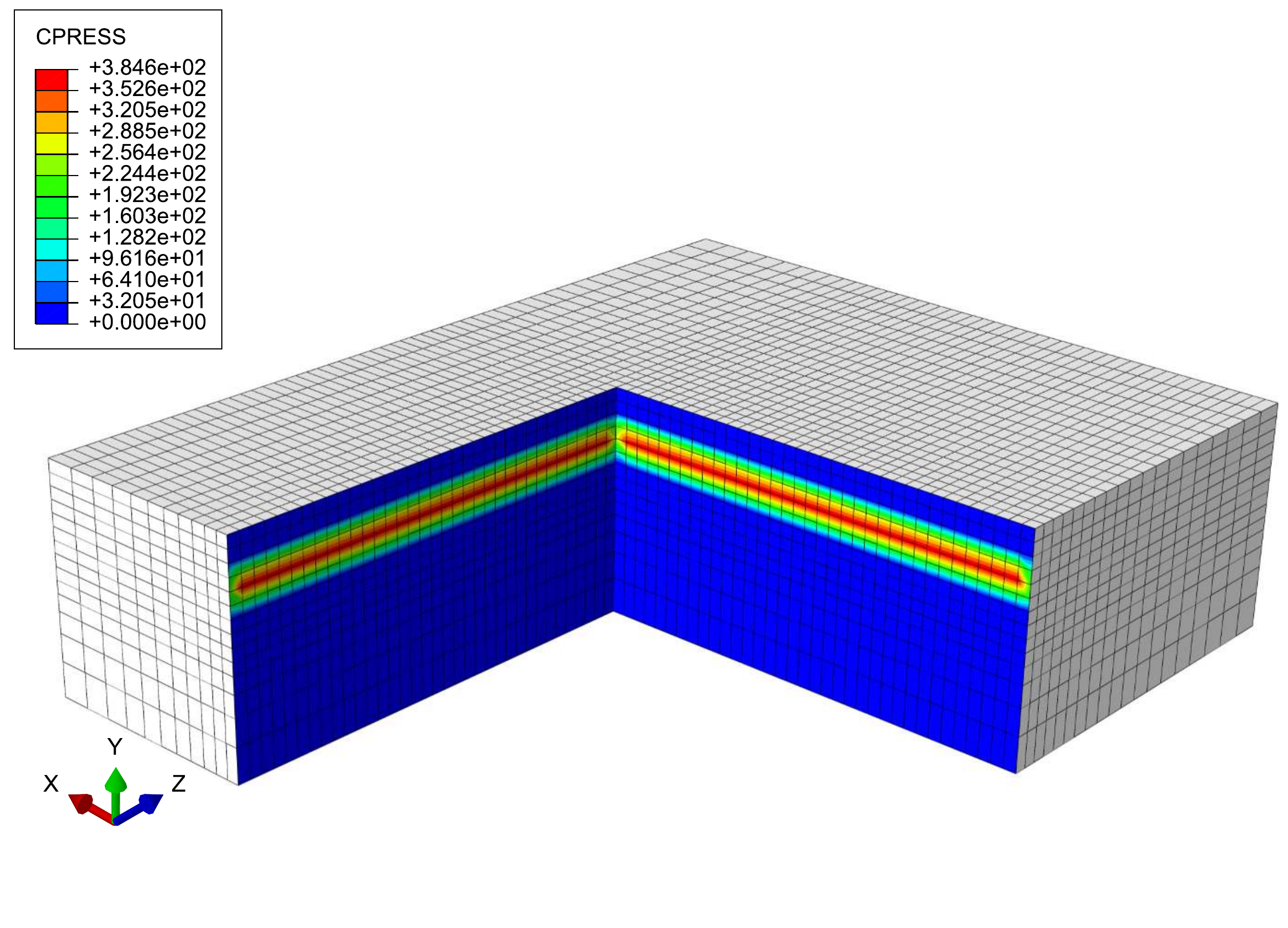}
\endminipage \\[6mm]
\minipage{0.48\textwidth}
  \includegraphics[width=\linewidth]{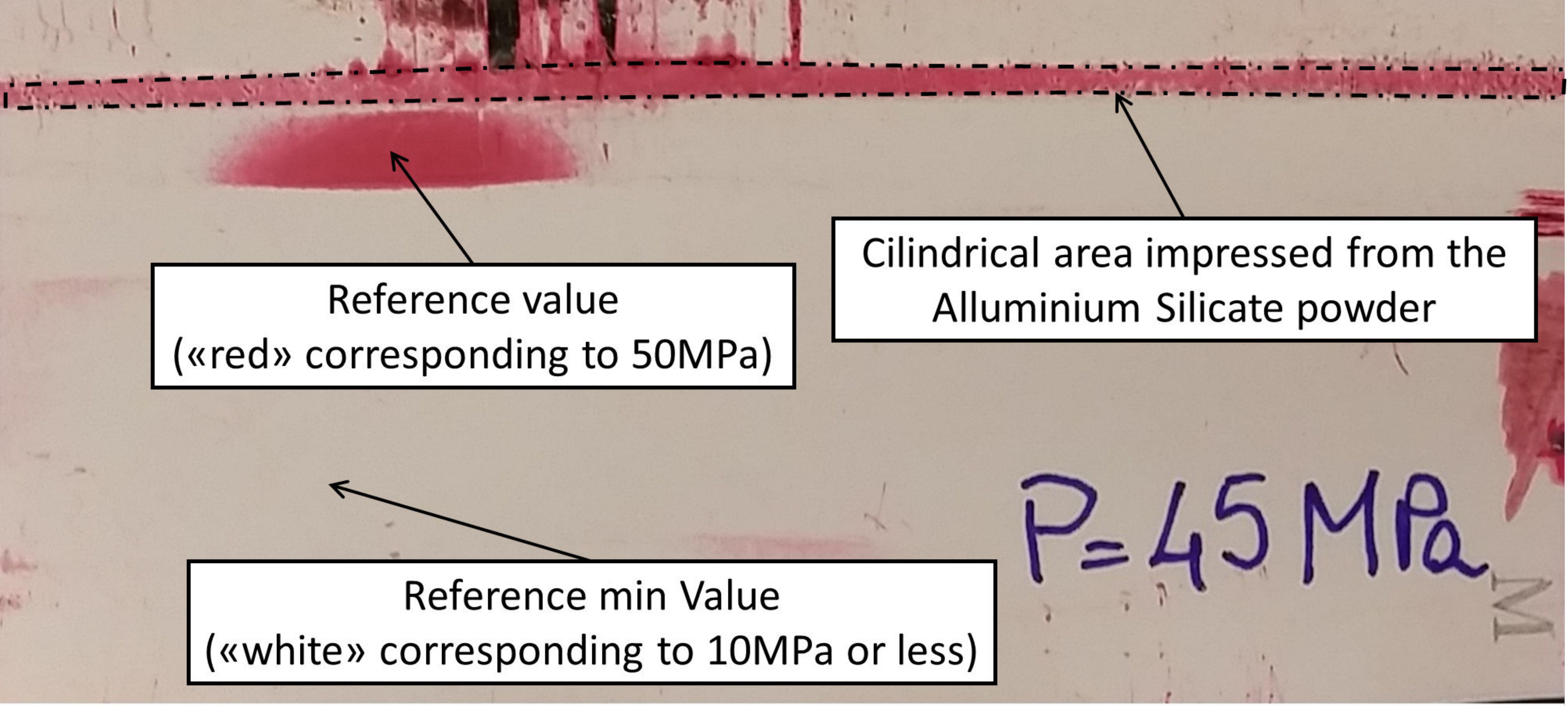}
\endminipage\hfil
\minipage{0.48\textwidth}
  \includegraphics[width=\linewidth]{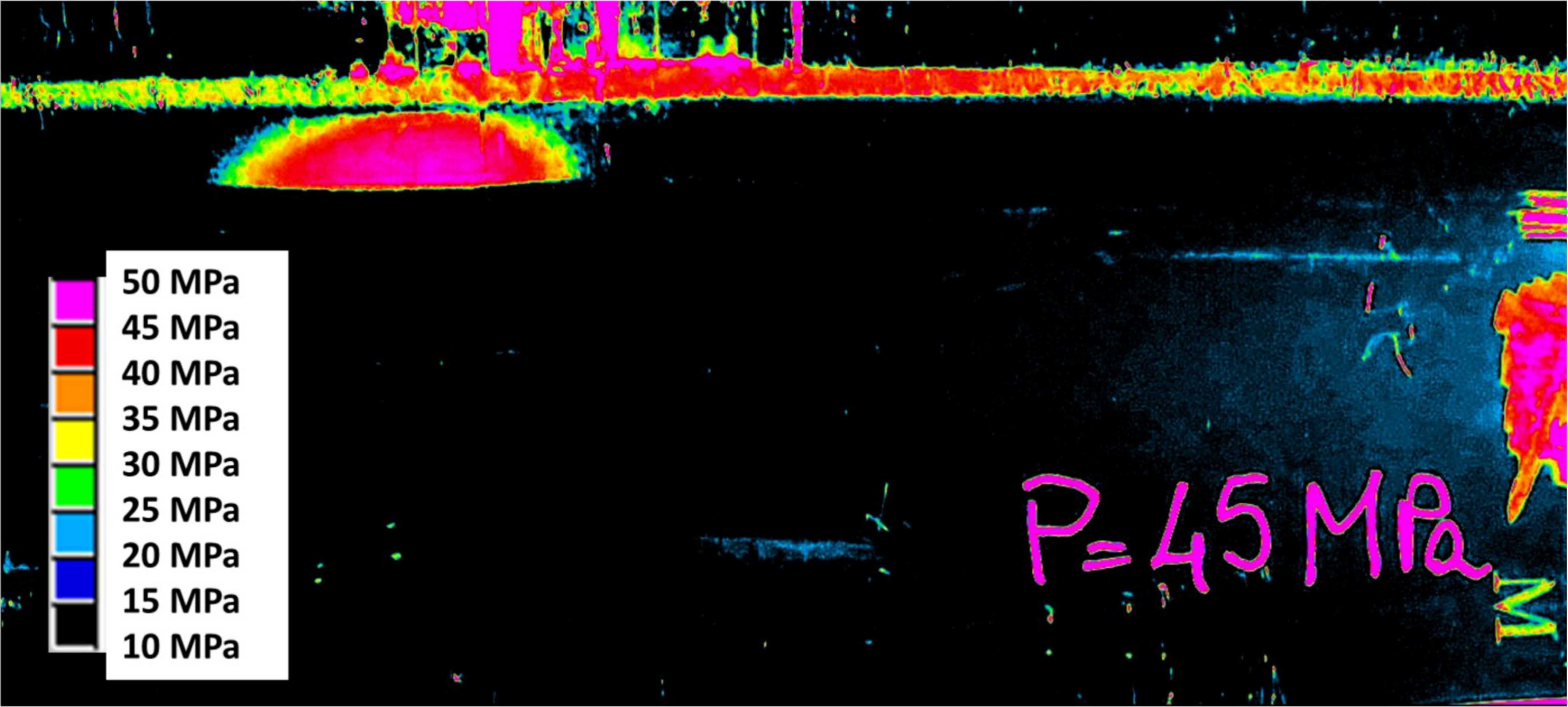}
\endminipage \\[6mm]
\caption{\footnotesize Contact pressure values in the three-dimensional simulation (upper part) and imprint left on the pressure-sensitive film after the experimental test (lower part).}
\label{pressure-sensitive-film}
\end{figure}

All FE analyses denoted high contact pressure values on the steel matrix. For an applied axial load of 45 MPa, three-dimensional and two-dimensional simulations yielded respectively average transversal pressure values equal to 37.73 MPa and to 39.45 MPa, respectively.
The contact pressure for the three-dimensional simulation is shown in Figure~\ref{pressure-sensitive-film} (upper part). The average contact pressure has been calculated dividing the resultant of the nodal contact forces by the contact area. 

These values of lateral contact pressure on the die wall are much higher than those usually considered in the design of tile forming devices. In the industrial practice, the lateral contact pressure is empirically assumed to be one sixth of the axial pressure applied on the powder.

\begin{figure}[!htcb]
\centering
\includegraphics[width=130mm]{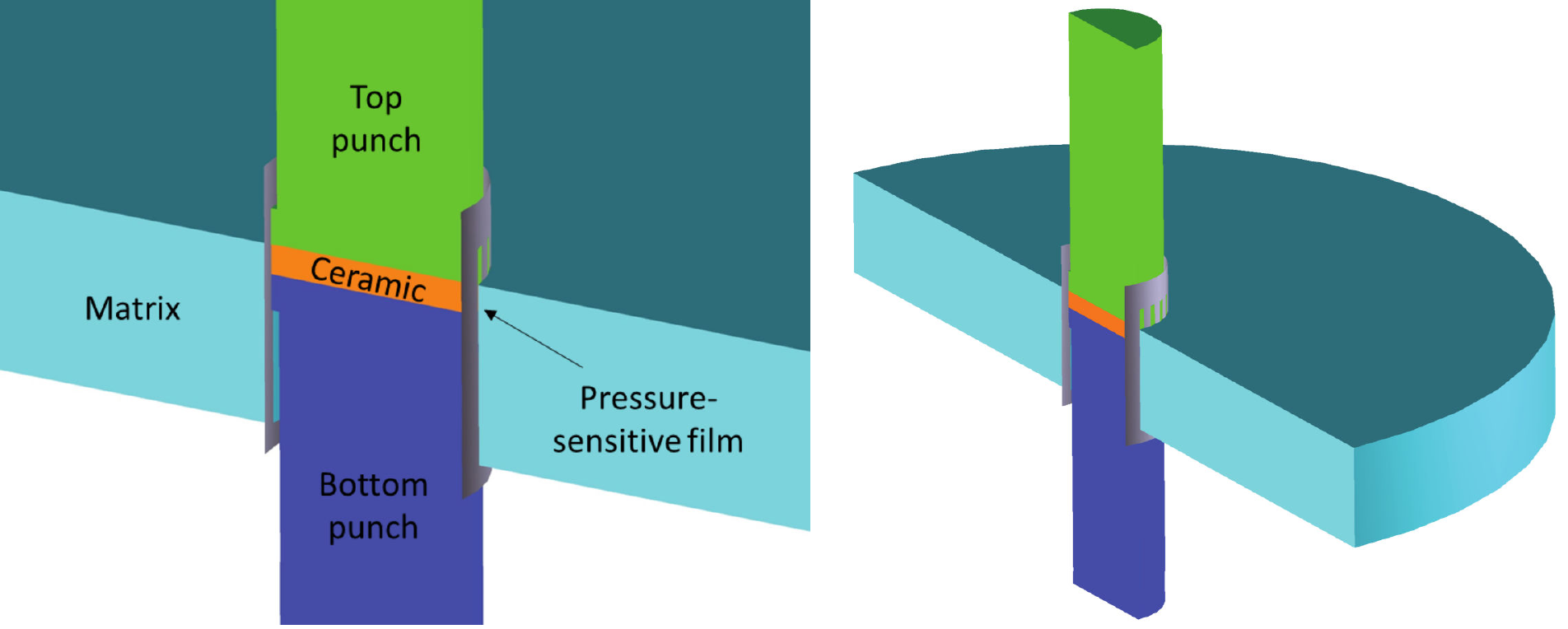}
\caption{\footnotesize Description of the performed experimental test: the pressure-sensitive film is placed between the ceramic powder and the steel matrix.}
\label{pressure-sensitive-film-scheme-together}
\end{figure}

To clarify this point, a specific experimental investigation has been performed, by placing a pressure-sensitive {\it Fuji Prescale MS} film between ceramic powder and steel matrix, see Fig.~\ref{pressure-sensitive-film-scheme-together}.

This film is composed of a polyester base on which a colour-developing material is coated, with the micro-encapsulated colour-forming material layered on top. When pressure is applied on the film, the microcapsules are broken and the colour-forming material reacts with the colour-developing material, and this process causes magenta colour forming. Microcapsules are designed to react to various degrees of pressures, releasing their colour forming material at a density that correspond to specific levels of applied pressure.

Figure~\ref{pressure-sensitive-film} (lower part) shows the imprint left by the ceramic powder on the pressure-sensitive film. 
The lateral pressure on the matrix was obtained by analysing the image using a dedicated program that convert magentascale to RGB values. For a forming axial pressure of 45 MPa, the measured average lateral pressure on the matrix is equal to 39 MPa, confirming the results of the simulations with a very good accuracy.

\clearpage

\subsection{Density distribution in a combed finish tile after die pressing}

\begin{figure}[!htcb]
\centering
\includegraphics[width=130mm]{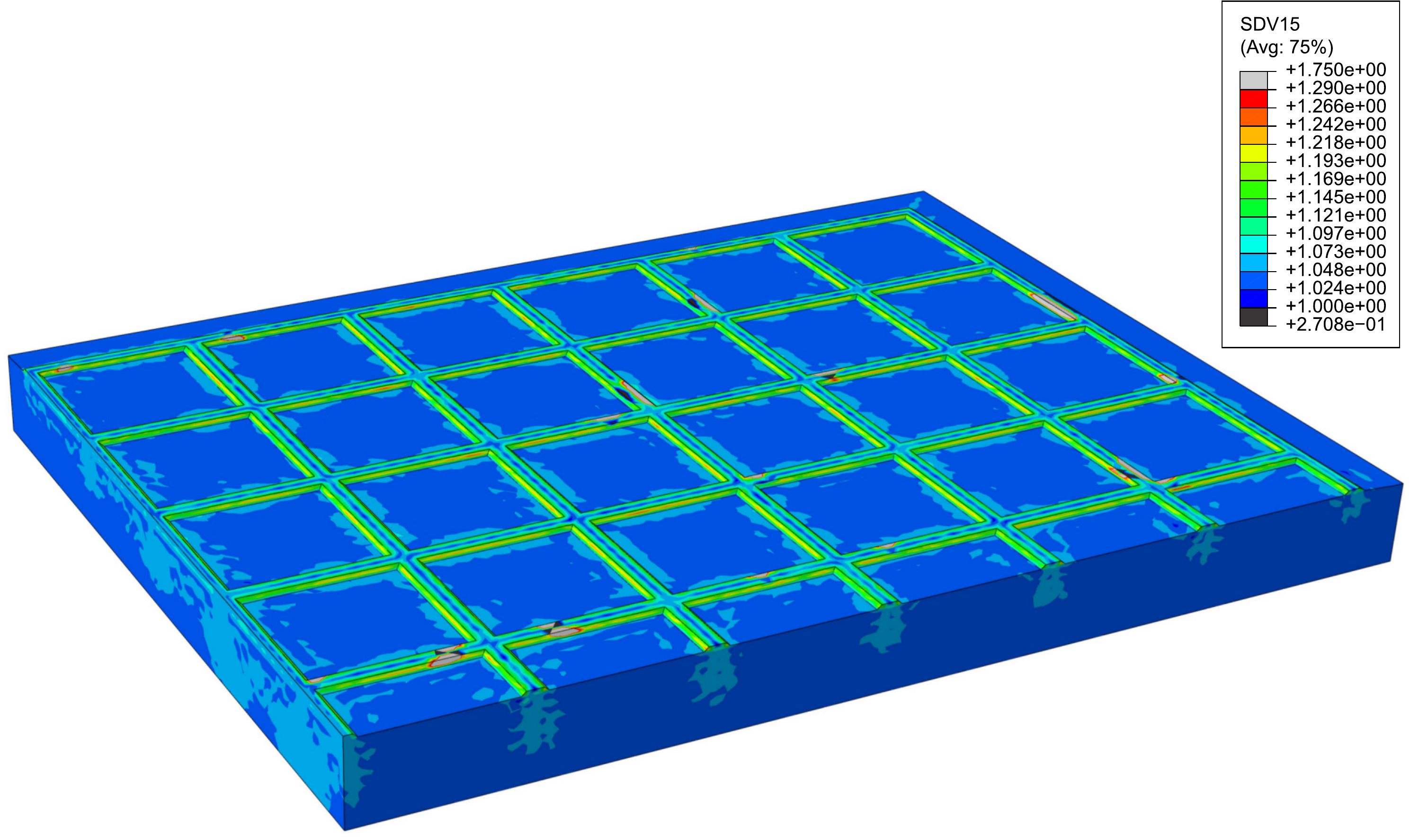}
\caption{\footnotesize Void ratio distribution in a combed finish tile after die pressing.}
\label{sdv_15_3cut}
\end{figure}

The density distribution in green bodies is of primary importance in the optimization of tile forming processes. In fact, non-uniform density can affect the subsequent sintering process, resulting in a poor quality of the final ceramic product. Figure~\ref{sdv_15_3cut} shows the void ratio distribution in a combed finish tile green body, predicted by the finite element simulation. The void ratio, defined as the ratio of the volume of voids to the volume of solid, $e = V_V/V_S$, is related to the relative density by
\begin{equation}
\label{voidratio}
\frac{\rho}{\rho_0} = \frac{1 + e_0}{1 + e},
\end{equation}
where $\rho_0$ and $e_0$ are the initial values.
We notice higher void ratio values, corresponding to lower density, in correspondence of the protrusions. These results pave the way to more advanced virtual prototyping analyses, aiming at the optimization of the design of tile forming devices, in order to produce green bodies with improved density distribution.

\section{Conclusions}

A simulation tool has been presented for the design and the optimization of a forming process of ceramic powders for the production of tiles. In the modelling composition, granulometry, morphology, and humidity of the powder is accounted for in a phenomenological  perspective, namely, through the introduction of constitutive parameters obtained from specifically-designed experimental tests. Other parameters needed for the analyses were shown to be deductible through a multi target optimization software. 
All these parameters allow for a refined description of the complex mechanical behaviour of the material occurring during cold forming. The software developed for the in-silico evaluation of the forming processes has been proven to be numerically robust and  has been validated through a comparison between numerical predictions and experimental measurements of the pressure to which a tile is exposed during forming. 
The availability of the simulation tool opens new possibilities in the design of manufacturing ceramic pieces and in the optimization of the production process to achieve higher production  quality requirements in terms of density distribution, residual stresses, dimension tolerance, cracking, and minimization of effects related to the spring-back.

\section*{Acknowledgements}

LA, MC, and DM acknowledge financial support from the European Union's Seventh Framework IAPP Programme PIAP-GA-2011-286110-INTERCER2.
MP and FDC acknowledge financial support from the European Union's Seventh Framework Programme PIAPP-GA-2013-609758-HOTBRICKS. 
AP acknowledges financial support from the European Union's Seventh Framework Programme PITN-GA-2013-606878-CERMAT2.

\bibliographystyle{jabbrv_kp}
\bibliography
{roaz}

%
%

\end{document}